\pdfoutput=1
\documentclass[aps,pre,preprint,nobibnotes,superscriptaddress,floatfix, 
showpacs,citeautoscript,showkeys]{revtex4-1}

\usepackage{graphicx}
\usepackage{bm}
\usepackage{amssymb}
\usepackage{amsbsy}
\usepackage{amsmath}
\usepackage{mathtools}
\usepackage{xcolor}
\usepackage{stmaryrd}
\usepackage{mathrsfs}
\usepackage{tikz}
\usetikzlibrary{shapes}
\usepackage{cancel}
\usepackage[normalem]{ulem}
\usepackage{hyperref}

\usepackage{cleveref}
\crefformat{section}{\S#2#1#3} 
\crefformat{subsection}{\S#2#1#3}
\crefformat{subsubsection}{\S#2#1#3}

\usepackage{color}
\usepackage{tikz}
\usetikzlibrary{shapes}
\usepackage{soul}

\newcommand\Pen{\mbox{\textit{Pe}}}

\usepackage{bm}

\DeclareBoldMathCommand{\bnabla}{\nabla}
\DeclareBoldMathCommand{\bcdot} {\cdot}
\DeclareBoldMathCommand{\btimes}{\times}

\makeatletter
\let\cat@comma@active\@empty
\makeatother

\makeatletter 
\gdef\@ptsize{2}
\let\@currsize\normalsize 
\makeatother 
\usepackage{setspace} 
\singlespace 

\DeclareRobustCommand{\markerfive}{\raisebox{0pt}{\tikz{\node[draw,scale=0.4,diamond,fill=none](){};}}}

\begin{document}

\title{Dynamics of forced and unforced autophoretic particles}

\author{R. Kailasham}
\affiliation{Department of Chemical Engineering, Carnegie Mellon University, Pittsburgh, PA 15213, USA}
\author{Aditya S. Khair}
\email{akhair@andrew.cmu.edu}
\affiliation{Department of Chemical Engineering, Carnegie Mellon University, Pittsburgh, PA 15213, USA}

\begin{abstract}
Chemically active, or autophoretic, particles that isotropically emit or absorb solute molecules undergo spontaneous self-propulsion when their activity is increased beyond a critical P\'{e}clet number ($\Pen$). Here, we conduct numerical computations, using a spectral-element based method, of a rigid, spherical autophoretic particle in unsteady rectilinear translation. The particle can be freely suspended (or `unforced') or subject to an external force field (or `forced'). The motion of an unforced particle progresses through four regimes as $\Pen$ is increased: quiescent, steady, stirring, and chaos. The particle is stationary in the quiescent regime, and the solute profile is isotropic about the particle. At $\Pen=4$ the fore-aft symmetry in the solute profile is broken, resulting in its steady self-propulsion. Our computations indicate that the self-propulsion speed scales linearly with $\Pen-4$ near the onset of self-propulsion, as has been predicted in previous studies. A further increase in $\Pen$ gives rise to the stirring regime at $\Pen\approx27$, where the fluid undergoes recirculation, while the particle remains essentially stationary. As $\Pen$ is increased even further, the particle dynamics are marked by chaotic oscillations at $\Pen\approx55$ and higher, which we characterize in terms of the mean square displacement and velocity autocorrelation of the particle. Our results for an autophoretic particle under a weak external force are in good agreement with recent asymptotic predictions (Saha, Yariv, and Schnitzer, \textit{J. Fluid Mech.}, vol. 916, A47, 2021). Additionally, we demonstrate that the strength and temporal scheduling of the external force may be tuned to modulate the chaotic dynamics at large $\Pen$.
\end{abstract}

\maketitle


\section{Introduction}
\label{sec:intro}

Autophoretic colloids that isotropically emit or absorb solute molecules at their surface are prototypical examples of synthetic active matter~\citep{Michelin2022b}. At low levels of chemical activity (quantified by a suitably small intrinsic P\'{e}clet number, \Pen), diffusion is dominant enough to homogenize perturbations to the solute distribution around such a particle, thereby rendering the concentration profile isotropic, and the particle stationary. Beyond a threshold P\'{e}clet number, however, small perturbations to the concentration field result in fore-aft symmetry breaking of the concentration profile, and the particle may undergo self-propulsion along a rectilinear path, or, at sufficiently large $\Pen$, execute meandering, helical and eventually chaotic motion. The spontaneous self-propulsion of autophoretic particles and droplets has been predicted theoretically~\citep{Michelin2013,Hu2019}, as well as being observed in experiments~\citep{Izri2014,Maass2016,Hokmabad2021,Suda2021} and numerical simulations~\citep{Michelin2013,Morozov2019,Hu2022}. 


The isotropic solute distribution around an autophoretic particle or drop that is free from an external force (i.e., an `unforced' particle)  is linearly unstable to dipolar concentration disturbances at $\Pen=4$~\citep{Michelin2013,Morozov2019}.
The fore-aft asymmetry of this disturbance leads to a net phoretic force on the particle, which is thereby set into spontaneous, steady self-propulsion in an arbitrary direction.
Weakly nonlinear analysis predicts that the speed at which the particle travels is asymptotic to $(\Pen-4)/16$ as $\Pen\to 4$~\citep{Morozov2019,Saha2021,Schnitzer2022}.
This prediction is in agreement with calculations by~\citet{Morozov2019}, who solved for the axisymmetric flow and solute fields around an autophoretic drop, via representation of the velocity field through the Stokes streamfunction, and the solute concentration as a series of Legendre polynomials.
By construction, the drop is constrained to unsteady, rectilinear translation.
Diffusiophoretic and Marangoni effects were considered as self-propulsion mechanisms; the relative importance of the former over the latter is represented by the scalar parameter $m$, with $m\to\infty$ and $m=1$ corresponding to purely diffusiophoretic and Marangoni forcing, respectively. 
The particle undergoes chaotic self-propulsion beyond a critical value of $\Pen$ (that increases with decreasing $m$), wherein its velocity varies erratically and its mean square displacement (MSD) undergoes long-time ballistic motion, scaling quadratically with respect to the lag time $\tau$.
These authors comment, however, that the observed behavior of the MSD may be due to their numerical solution scheme.

Very recently,~\citet{Hu2022} investigated the various trajectories undertaken by an unforced autophoretic particle as the P\'{e}clet number is increased, using both axisymmetric and fully three dimensional (3D) models for the particle motion.
In the latter case, the velocity field is represented by Lamb's solution to the Stokes equations, and the solute concentration is expanded in spherical harmonics.
Their axisymmetric computations indicate that the self-propulsion speed $U$ of the particle undergoes a regular pitchfork bifurcation near $Pe=4$, with the scaling given by $U\sim\left(Pe-4\right)^{1/2}$.
The same scaling was observed by~\citet{Li2022} near the onset of spontaneous motion of a two-dimensional (i.e., circular) autophoretic drop.
As mentioned above, other theoretical and numerical investigations, however, predict a singular pitchfork bifurcation at $Pe=4$, with a linear scaling of the self-propulsion speed, $U\sim\left(Pe-4\right)/16$~\citep{Morozov2019,Saha2021,Schnitzer2022}. 
Additionally, axisymmetric computations by~\citet{Hu2022} performed up to $\Pen=60$ indicate that the MSD displays short- and long-time ballistic scaling, with the velocity time series undergoing a period-doubling cascade toward the chaotic regime.
Their three-dimensional simulations, on the other hand, indicate that the particle undergoes long-time diffusion (i.e., MSD scaling linearly with $\tau$) following a ballistic regime at early-times, which is accompanied by an intermittency scenario in the chaotic velocity time series.
 
In the present paper, we first chart the transition of an unforced rigid, spherical autophoretic particle from steady self-propulsion to chaotic motion, using a spectral element based algorithm to  solve the unsteady advection-diffusion equation for the solute concentration, and the Stokes equations governing the quasi-steady velocity field.
Our computations are performed on an axisymmetric model, so the motion is restricted to unsteady rectilinear translation.
We find that the self-propulsion speed $U$ is well described by the asymptotic result $U\sim\left(Pe-4\right)/16$ near the onset of spontaneous motion. 
At $\Pen=55$ and $58$, i.e., within the period-doubling cascade identified by~\citet{Hu2022}, we do not observe a clear transition to long-time ballistic motion within the duration of $\tau$ computed.
Our computations at $\Pen=80$ indicate that the long-time behavior of the MSD is closer to diffusive than ballistic, in line with the 3D simulations by~\citet{Hu2022}.
Notably, the chaotic velocity times series at $\Pen=80$ shows intermittent characteristics, again in agreement with the 3D simulations by~\citet{Hu2022}.
Intermittency is absent for our computations at $\Pen=55$ and $58$, which appear to show period doubling behavior.
Despite the differing characteristics of the velocity time series from $\Pen=55$ to $\Pen=80$, the velocity autocorrelation (VAC) across this range of $\Pen$ varies smoothly, with the characteristic time for decorrelation decreasing with an increase in the P\'{e}clet number.

The second goal of this paper is to present computations on an autophoretic particle under an external force field, i.e. a `forced' particle.
The dynamics of a forced particle under a weak, steady external force was recently analyzed by~\citet{Saha2021}, who derived asymptotic relations for the self-propulsion speed near the bifurcation point, $\Pen=4$, of the unforced problem.
Here, the imposition of an external force leads to an imperfect bifurcation~\citep{Michelin2022b}, from a perturbed stationary state to a perturbed spontaneous motion.
Remarkably, they predicted that the latter could occur parallel or anti-parallel to the direction of the external force.
We start by validating our numerical computations against their analytical approximations. 
After that, we quantify the effect of an external force on particle dynamics in the chaotic regime, including the effect of a temporal modulation in the force. The dynamics of autophoretic particles under an external force or a flow field has been discussed recently in a review by~\citet{Michelin2022b}, which also examines the effect of such external forcing on the stability of the different branches in the bifurcation diagram.

The rest of the paper is organised as follows. In~\cref{sec:prob_form}, we specify the governing equations for the dynamics of a rigid, spherical autophoretic particle. 
The numerical details of the solution to the coupled nonlinear partial differential equations governing the transient evolution of the concentration field and the quasi-steady velocity field are provided in~\cref{sec:num_details}, along with the details of the MSD and the VAC calculations. 
We present and discuss the results for the unforced particle in~\cref{sec:ep_zero_results}, followed by that for the forced particle in~\cref{sec:ep_fin_results}. 
A conclusion is offered in~\cref{sec:conclusion}.

\section {Problem formulation}
\label{sec:prob_form}
We consider a rigid, spherical autophoretic particle of radius $a^{*}$ in an incompressible Newtonian fluid of viscosity $\eta^{*}$, whose flow obeys the Stokes equations. Above and henceforth, dimensional variables are marked with an asterisk as superscript. There is a constant flux $\mathcal{A}^{*}$ of solute at the particle surface, which is positive (negative)  when the solute is being emitted (absorbed) at the interface. Far away from the particle, the uniform solute concentration is $C^{*}_{\infty}$, and the difference between the local concentration and its far-field value is denoted by $c^{*}=C^{*}-C^{*}_{\infty}$. The solute molecules, of diffusivity $D^{*}$, interact with the active particle via a short-ranged potential whose characteristic length is $b^{*}$, such that $b^{*}\ll a^{*}$. These interactions give rise to a tangential slip velocity along the particle surface~\citep{Anderson1989}, whose magnitude is set by the concentration gradient of the solute at the interface, and the mobility parameter, $\mathcal{M}^{*}=\pm {k_B^{*}T^{*}b^{*2}}/{\eta^{*}}$, where $k^{*}_{B}$ is Boltzmann's constant and $T^{*}$ is the absolute temperature. Attractive (repulsive) interactions between the solute and the active particle are described by a negative (positive) $\mathcal{M}^{*}$~\citep{Michelin2014}. The concentration profile of the solute, therefore, evolves due to diffusion and advection by fluid flow. 

Following~\citet{Michelin2013}, length, time, fluid velocity, pressure, and concentration are scaled by $a^{*}$, ${a^{*}D^{*}}/{|\mathcal{A}^{*}\mathcal{M}^{*}|}$, $U^*={|\mathcal{A}^{*}\mathcal{M}^{*}|}/{D^{*}}$, ${\eta^{*}U^*}/{a^{*}}$, and ${a^{*}|\mathcal{A}^{*}|}/{D^{*}}$, respectively. 
Additionally, following~\citet{Saha2021} an external force of magnitude $F^*$ imposed on the particle leads to a `mechanical' velocity scale $W^*=F^*/(6\pi\eta^*a^*)$.
In the subsequent discussion, variables without asterisks are the dimensionless equivalent of their dimensional counterparts. A key dimensionless parameter is the intrinsic P\'{e}clet number,
\begin{equation}
\Pen=\dfrac{a^{*}|\mathcal{A}^{*}\mathcal{M}^{*}|}{D^{*2}},
\end{equation}
which quantifies the relative importance of the solute advection with respect to its diffusion, and is a measure of the chemical activity of the particle. 
Another important parameter is the ratio of `mechanical' to `chemical' velocity scales, $\epsilon=W^*/U^*$, which is by assumption small compared to unity in the work of~\citet{Saha2021}.
It is convenient to introduce the scaled flux and mobility parameters, 
\begin{equation}
A=\dfrac{\mathcal{A}^{*}}{|\mathcal{A}^{*}|},\,M=\dfrac{\mathcal{M}^{*}}{|\mathcal{M}^{*}|}.
\end{equation}

The concentration field is governed by the unsteady advection-diffusion equation,
\begin{equation}\label{eq:adv_diff}
\Pen\left(\dfrac{\partial c}{\partial t}+\boldsymbol{v}\bcdot\bnabla c\right)=\nabla^2 c,
\end{equation}
where $t$ is time and $\boldsymbol{v}$ denotes the velocity field. Eq.~\ref{eq:adv_diff} is subject to the two boundary conditions of: (i) constant flux of the solute at the surface of the active particle
\begin{equation}\label{eq:const_flux}
\dfrac{\partial c}{\partial r}=-A\quad\text{at}\quad r=1,
\end{equation}
and (ii) an attenuation condition far away from the particle
\begin{equation}\label{eq:atten}
c\to0\quad\text{as}\quad r\to\infty.
\end{equation}

We employ a cylindrical coordinate system $(z,\rho,\phi)$ with its origin attached to the particle centre. The $z$-axis represents the axis of symmetry along which the particle motion is constrained. The perpendicular distance from the $z$-axis is measured by the $\rho$-coordinate, and $\phi$ denotes the azimuthal angle of rotation about the $z$-axis. This frame of reference is non-inertial since the particle is accelerating during its unsteady translation. However, this choice does not affect the advection-diffusion equation or the fluid flow equations, since the motion occurs at zero Reynolds number. The particle surface is denoted by $r=1$ where $r=\sqrt{\rho^2+z^2}$. The axisymmetric flow around the particle is represented as
\begin{equation}
\boldsymbol{v}={v}_{z}(\rho,z)\boldsymbol{e}_{z}+{v}_{\rho}(\rho,z)\boldsymbol{e}_{\rho},
\end{equation}
where $\boldsymbol{e}_{z}$ and $\boldsymbol{e}_{\rho}$ are unit vectors along the $z$ and $\rho$ axis, respectively. It is useful to define the polar angle $\theta\equiv\arctan(\rho/z)$ measured anticlockwise such that $\theta=0$ lies on the positive $z$-axis.

The velocity field, $\boldsymbol{v}$, in (\ref{eq:adv_diff}) is governed by the incompressibility criterion and the Stokes equation,
\begin{equation}\label{eq:stokes_raw}
\bnabla\bcdot\boldsymbol{v}=0,\nabla^2\boldsymbol{v}=\bnabla p,
\end{equation}
where $p$ is the dynamic pressure, and subject to the following boundary conditions
\begin{equation}\label{eq:slip_con}
\boldsymbol{v}\equiv M\bnabla_{\text{s}}c=v_{\text{s}}\left(\cos\theta\boldsymbol{e}_{\rho}-\sin\theta\boldsymbol{e}_{z}\right)\quad\text{at}\quad r=1,
\end{equation}
where
\begin{equation}
v_{\text{s}}=M\left(\cos\theta\dfrac{\partial c}{\partial \rho}-\sin\theta\dfrac{\partial c}{\partial z}\right),
\end{equation}
and
\begin{equation}\label{eq:far_field_vel}
\boldsymbol{v}\to-U\boldsymbol{e}_{z}\quad\text{as}\quad r\to\infty.
\end{equation}
The far-field speed $U$ in eq.~(\ref{eq:far_field_vel}) is unknown \textit{a priori}, and is determined by requiring that the total hydrodynamic force on the particle in the $z$-direction is at all times is equal to $-6\pi\epsilon$. The presumed axisymmetry of the problem permits us to rewrite the velocity field in terms of the Stokes streamfunction $\psi$,
\begin{equation}\label{eq:vel_def}
\boldsymbol{v}=\dfrac{1}{\rho}\left(\boldsymbol{e}_{z}\dfrac{\partial \psi}{\partial \rho}-\boldsymbol{e}_{\rho}\dfrac{\partial \psi}{\partial z}\right)\equiv \dfrac{1}{\rho}{\bnabla}^{\perp}\psi.
\end{equation}
Taking the curl of the Stokes equations and introducing the vorticity vector $\boldsymbol{\omega}=\bnabla\times\boldsymbol{v}$ eliminates the pressure from the governing equations, resulting in the following system of coupled partial differential equations,
\begin{equation}\label{eq:vort_def}
\omega\rho+E^2\psi=0,
\end{equation}
\begin{equation}\label{eq:vort_transport}
\nabla^2 \omega-\dfrac{\omega}{\rho^2}=0,
\end{equation}
where $\omega$ is the $\phi$-component of the vorticity $\boldsymbol{\omega}$ about the $z$-axis, with the other components ($\rho$ and $z$) of the vorticity vector vanishing due to symmetry, and the operators
\begin{equation}\label{eq:grad_op_def}
\nabla^2=\dfrac{1}{\rho}\dfrac{\partial}{\partial \rho}\left(\rho\dfrac{\partial}{\partial \rho}\right)+\dfrac{\partial^2}{\partial z^2},\,E^2=\nabla^2-\dfrac{2}{\rho}\dfrac{\partial}{\partial \rho}.
\end{equation}
Equations~(\ref{eq:vort_def}) and ~(\ref{eq:vort_transport}) must be solved subject to the following boundary conditions
\begin{equation}\label{eq:dn_grad_sfn_near}
\boldsymbol{n}\bcdot\bnabla\psi=-\rho v_{\text{s}}\quad\text{at}\quad r=1,
\end{equation}
\begin{equation}\label{eq:dn_grad_sfn_far}
\boldsymbol{n}\bcdot\bnabla\psi\to-\rho U\sin\theta\quad\text{as}\quad r\to\infty,
\end{equation}
\begin{equation}\label{eq:dn_grad_vort_far}
\boldsymbol{n}\bcdot\bnabla\omega\to0\quad\text{as}\quad r\to\infty,
\end{equation}
\begin{equation}
\psi=0\quad\text{at}\quad r=1,
\end{equation}
with $\boldsymbol{n}=\cos\theta\boldsymbol{e}_{z}+\sin\theta\boldsymbol{e}_{\rho}$ being the unit normal to the particle surface pointing into the fluid;
and the symmetry condition
\begin{equation}\label{eq:symaxis}
\psi=\omega=0\quad\text{along}\quad \rho=0.
\end{equation}
 
For an unforced particle ($\epsilon=0$) there exists a trivial solution to the system of equations (\ref{eq:adv_diff}), (\ref{eq:vort_def}) and (\ref{eq:vort_transport}), which is the isotropic concentration profile $c=1/r$, corresponding to the quiescent state of zero phoretic velocity and no fluid motion at all times. Beyond $\Pen=4$, however, the quiescent state becomes unstable with respect to dipolar perturbations in the concentration field~\citep{Michelin2013,Morozov2019}, and the autophoretic particle sets into motion. Further types of perturbation (such as quadrupolar) are unstable at larger values of the P\'{e}clet number as the higher hydrodynamic modes are excited. We therefore supply an asymmetrical concentration field as the initial condition to eq.~(\ref{eq:adv_diff}), as follows
\begin{equation}\label{eq:init_cond}
c(r,t=0)=\dfrac{1}{r}-\delta_{\text{per}}\left(\dfrac{\cos\theta}{r^2}\right),
\end{equation}
with $|\delta_{\text{per}}|<1$.

{For a forced particle ($\epsilon\neq 0$), there exists no trivial solution corresponding to an isotropic concentration profile. Nonetheless, we still use an initial condition of the form given by eq.~(\ref{eq:init_cond}). The agreement of the resultant numerical results with the asymptotic predictions of~\citet{Saha2021} (discussed in~\cref{sec:ep_fin_results}) justifies the usage of this initial condition.}

\section {Numerical solution methodology}
\label{sec:num_details}

Numerical computations require the stipulation of a finite outer boundary, and a value of $R_{\text{o}}=100$ is chosen as the radius of the spherical shell on which the far-field boundary conditions are prescribed, unless specified otherwise. {The far-field boundary conditions, eqs. (2.5) and (2.10), are enforced exactly at $r=R_{\text{o}}$. The computational domain is a polar grid in the $(r,\theta)$ space, with $N_{r}=N_{\theta}=15$ quadrilateral elements in both the radial and angular directions. The elements in the $\theta-$ direction are evenly spaced in the range $[0,\pi]$, while those in the $r$-direction are arranged such that the size of the radial elements follow a geometrical progression, with the width of the radial element closest to the particle surface being $\Delta r_{0}=0.388$.}

As the first step in the numerical solution process, the initial condition given in~(\ref{eq:init_cond}) is used to specify the slip boundary condition~(\ref{eq:slip_con}) for the flow problem. The self-propulsion speed $U(t)$ at any instant of time $t$ must satisfy the requirement that the total hydrodynamic drag force on the particle in the $z$-direction, $F_{z}$, given by~\citep{Khair2014}
\begin{equation}\label{eq:drag_calc}
F_{z}=\left\{\pi\int_{0}^{\pi}\left[\dfrac{\partial(\omega r)}{\partial r}-2\omega\right]_{r=1}\sin^2\theta\,d\theta\right\}+6\pi\epsilon
\end{equation}
vanishes at $t$. The self-propulsion speed is evaluated iteratively using a secant method, as described in~\cite{Chisholm2016}. Given two initial guesses $U^{\left<\ell\right>}$ and $U^{\left<\ell-1\right>}$ at time $t$, where $\ell$ denotes the iteration number, the hydrodynamic drag at the two values of the self-propulsion speed are evaluated. Using linear interpolation, an improved estimate for $U$ is obtained as: $U^{\left<\ell+1\right>}=\left(U^{\left<\ell\right>}F_z^{\left<\ell-1\right>}-U^{\left<\ell-1\right>}F_z^{\left<\ell\right>}\right)/\left(F_z^{\left<\ell-1\right>}-F_z^{\left<\ell\right>}\right)$. The procedure is terminated when the magnitude of the difference in the computed speeds between successive iterations $|U^{\left<\ell\right>}-U^{\left<\ell-1\right>}|$ is reduced below $10^{-5}$. The converged solution for the flow field $\left(\psi,\omega\right)$ at each time instant is then used to solve the advection-diffusion equation for the concentration field. We present next the algorithm for the solution of the advection-diffusion equation governing the transient evolution of concentration (eq.~(\ref{eq:adv_diff})). The Stokes equations governing the flow field are solved in the same manner as in~\cite{Chisholm2016} and ~\cite{Khair2018}.

The time-derivative in eq.~(\ref{eq:adv_diff}) is discretized using the finite-difference formula\begin{equation}\label{eq:euler_def}
\dfrac{\partial c}{\partial t}\approx \dfrac{c^{(n+1)}-c^{(n)}}{\Delta t},
\end{equation}
where $c^{(n)}$ denotes the value of the concentration at the discrete time $t^{(i)}$, and $\Delta t\equiv t^{(n+1)}-t^{(n)}$ is the width of the discrete timestep. 

The concentration variable is taken to be the weighted sum of its values at its current and previous timestep, that is, 
\begin{equation}\label{eq:crank_nicol}
c=\Theta c^{(n+1)}+\left(1-\Theta\right)c^{(n)},
\end{equation}
where the choice of the $\Theta$ parameter corresponds to different well-known methods for time-discretization. For example, $\Theta=1$ represents the backward Euler method, while $\Theta=1/2$ is used in the Crank-Nicolson algorithm. We have used the latter method throughout this paper, in view of its unconditional stability and second-order accuracy~\citep{Tanaka1994,DoneaHuerta2003} with respect to the discretization width, $\Delta t$. A value of $\Delta t=1.0$ is used for simulations with $\Pen<20$, while a timestep width of 0.1 is used at higher values of the P\'{e}clet number. In the discussion that follows, we have retained the $\Theta$ notation for the sake of generality.

Substituting Eq.~(\ref{eq:euler_def}) into Eq.~(\ref{eq:adv_diff}), the governing equation may be rewritten as
\begin{equation}\label{eq:sem-disc_Theta_form}
\begin{split}
&-\left(\Theta\Delta t\right)\nabla^2c^{(n+1)}+{\Pen}\,c^{(n+1)}+\text{\Pen}\left(\Theta\Delta t\right)\left(\boldsymbol{v}\cdot{\bnabla}c^{(n+1)}\right)\\
&-\left(1-\Theta\right)\Delta t\nabla^2c^{(n)}-{\Pen}\,c^{(n)}+\text{\Pen}\Delta t\left(1-\Theta\right)\left(\boldsymbol{v}\cdot{\bnabla}c^{(n)}\right)=0
\end{split}
\end{equation}
where the velocity $\boldsymbol{v}$ is known from the solution of eqs.~(\ref{eq:vort_def}) and~(\ref{eq:vort_transport}). The weak variational form of Eq.~(\ref{eq:sem-disc_Theta_form}) is obtained by multiplying each term by a test function and taking an inner product~\citep{Khair2018,CampionRenson1978}. The concentration field is discretized, and the independent variables ($\rho,z$) parametrized using a set of shape functions defined as tensor products of 1D Lagrange polynomials of high order ($N_{\text{o}}=8$), supported at $N_{\text{o}}+1$ Gauss-Lobatto quadrature points over the standard region $[-1,1]^2$. A global matrix equation over all spectral elements is then assembled, and solved iteratively until the $L^2$-norm of the difference in concentration between successive iterations computed over all discretization points at each timestep drops below $10^{-6}$. Convergence studies with respect to the outer shell radius ($R_{o}$) and the integration timestep ($\Delta t$) are provided in Appendix~\ref{sec:app_a}.

Previous investigations~\citep{Michelin2013,Hu2019} on autophoresis have established that for unforced particles ($\epsilon=0$) with oppositely signed $A$ and $M$, perturbations to the concentration field vanish in the long-time limit, where the particle remains stationary. We therefore concern ourselves with the nontrivial case of similarly signed $A$ and $M$ values, which result in the self-propulsion of the particle, and we pick $A=M=1$ without loss of generality.
We examine both  $AM=1$ and $AM=-1$ for a forced particle.

At $\Pen\leq10$ the solver is supplied an initial condition of the form given by eq.~(\ref{eq:init_cond}), with the choice of $\delta_{\text{per}}=0.1$. It is observed that different values of $\delta_{\text{per}}$ in this regime result in the same long-time prediction for the steady phoretic velocity. At $\Pen>10$ the concentration profile computed at a previous (lower) value of the P\'{e}clet number is used as the initial condition for the subsequent simulation at the next higher \Pen.

The transient solver for the concentration field described above may be used across a range of $\Pen$, because it does not make any assumptions about the steadiness of the self-propulsion, and thereby allows for the phoretic velocity and the concentration field to be functions of time. An alternative solution methodology, valid (and more efficient) at low values of $\Pen$, where the particle is either stationary or undergoing steady translation, is an iterative technique, which solves the coupled system of equations given by the steady advection-diffusion equation,
\begin{equation}
\Pen\left(\boldsymbol{v}\bcdot\bnabla c\right)=\nabla^2 c
\end{equation}
and eq.~(\ref{eq:stokes_raw}), using the spectral element method, and subject to the same boundary conditions as described previously. The calculation is started by providing a concentration field of the form given by the RHS of eq.~(\ref{eq:init_cond}) as the initial guess, and repeated until the difference between the phoretic velocities obtained in successive iterations differ by less than $10^{-5}$. The utility of the iterative solver is admittedly limited in comparison to the transient solver. Nonetheless, the results obtained using the iterative solver prove useful in comparison against analytical approximations for the self-propulsion speed of unforced and forced autophoretic particles near $\Pen=4$, as discussed below in connection with figures~\ref{fig:steady_swim} and ~\ref{fig:comp_schnitz}. All results presented henceforth in this paper have been obtained using the transient solver, unless mentioned otherwise.

The axisymmetry of the flow field restricts motion of the particle along the $z$ axis. From the computed time-series of the self-propulsion speed, the particle position at a discrete time $i+1$ is evaluated as 
\begin{equation}\label{eq:z_pos}
z({i+1})=z({i})+U(i)\Delta t;\,\,z(0)=0.
\end{equation}
In the time interval $[0,t_{\text{sim}}]$, a total of $N\equiv t_{\text{sim}}/\Delta t$ data points are recorded. The MSD and VAC, $C_v$, at the $n^{\text{th}}$ time interval are evaluated as follows
\begin{equation}\label{eq:msd_def}
\text{MSD}(n)=\dfrac{1}{N-n}\sum_{i=0}^{N-n}\left[z(i+n)-z(i)\right]^2,
\end{equation}
\begin{equation}\label{eq:vac_def}
C_{v}(n)=\dfrac{1}{N-n}\sum_{i=0}^{N-n}\left[\boldsymbol{e}_{u}(i+n)\bcdot\boldsymbol{e}_{u}(i)\right],
\end{equation}
where $\boldsymbol{e}_{u}(i)\equiv U(i)/|U(i)|$ is the unit direction vector of the self-propulsion speed at discrete time $i$, and can take the value of $+1$ or $-1$. Equations~(\ref{eq:msd_def}) and ~(\ref{eq:vac_def}) represent calculations of the time average over a single trajectory. The lag-time is defined as $\tau\equiv n\Delta t$ and used in reporting the MSD and VAC results. For $\Pen\geq50$, the total simulation time is at least $t_{\text{sim}}=5\times10^3$, while that for $\Pen=75$ and $\Pen=80$ are $t_{\text{sim}}=1\times10^4$ and $t_{\text{sim}}=2\times10^4$, respectively. Data points corresponding to at least the first $1\times 10^4$ timesteps in these runs are discarded prior to the calculation of the MSD and VAC, to remove the effects of transients. 

\section {Unforced autophoretic particle}
\label{sec:ep_zero_results}

\begin{figure}[h]
  \centerline{\includegraphics[width=4.8in,height=!]{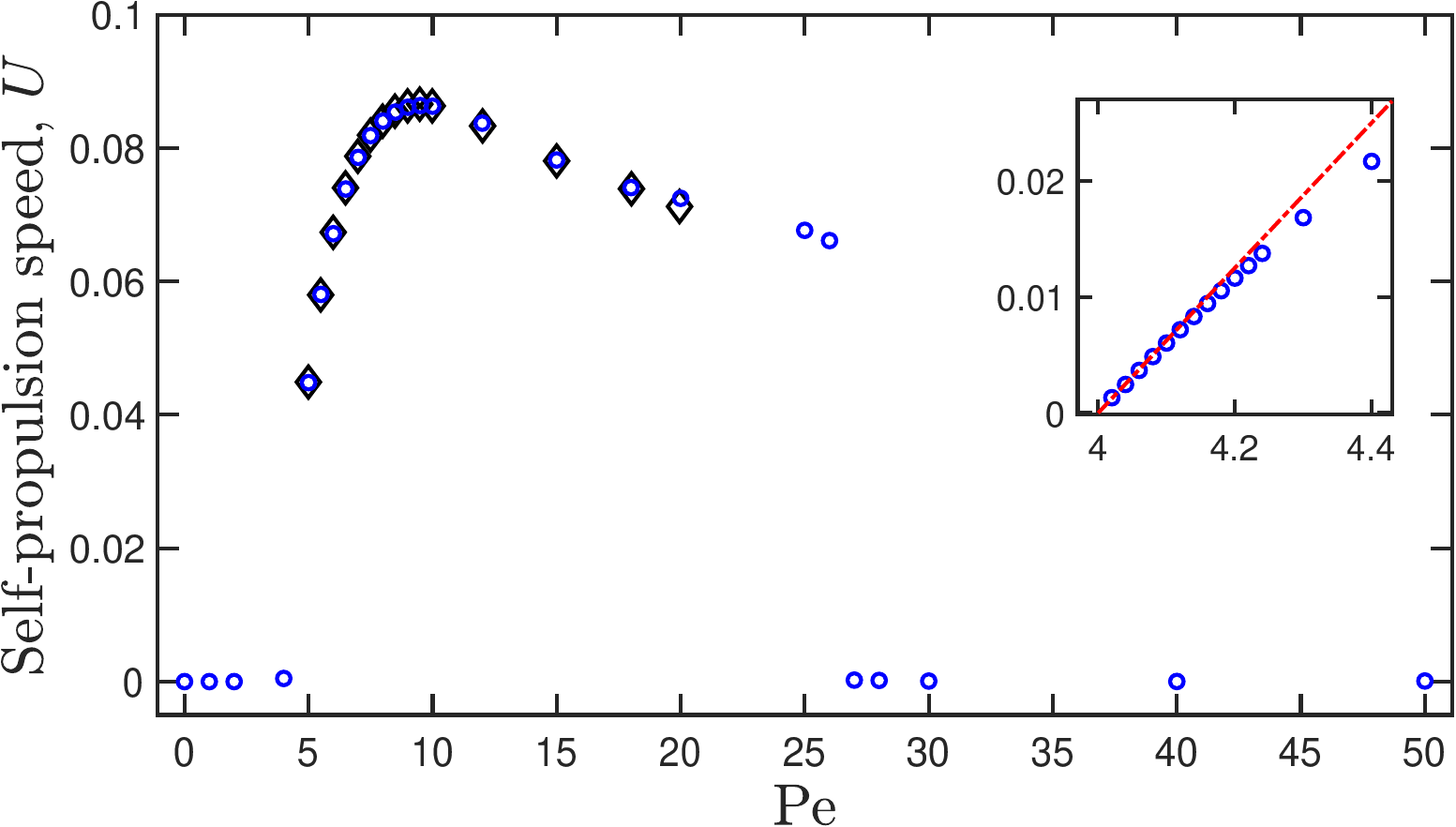}}
  \caption{ (Colour online) Steady self-propulsion speed of an unforced autophoretic particle at various P\'{e}clet numbers across the quiescent, steady, and stirring regimes. The unfilled diamonds (\markerfive) represent results obtained by~\citet{Michelin2013}, and the unfilled circles are from the present work. The inset compares the numerical results obtained using the iterative solver with $R_o=1000$ in the vicinity of the first bifurcation against the analytical approximation, $U=\left(\Pen-4\right)/16$ (indicated by line), derived by~\citet{Morozov2019} and ~\citet{Saha2021}.}
\label{fig:steady_swim}
\end{figure}

\begin{figure}
\begin{center}
\begin{tabular}{cc}
\includegraphics[width=2.5in, scale=0.8]{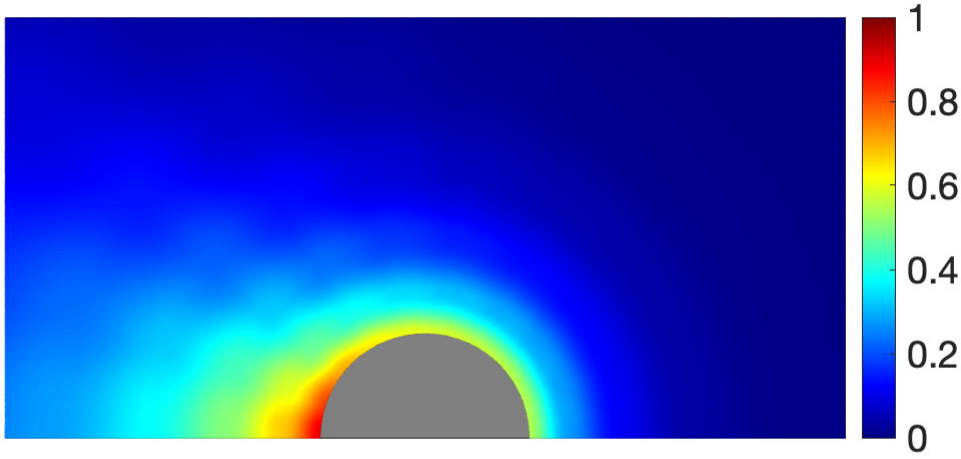}&
\includegraphics[width=2in, scale=0.8]{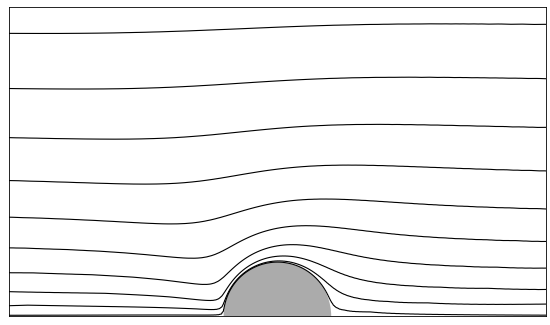}\\[5pt]
(a)&(b)\\
\includegraphics[width=2.5in, scale=0.8]{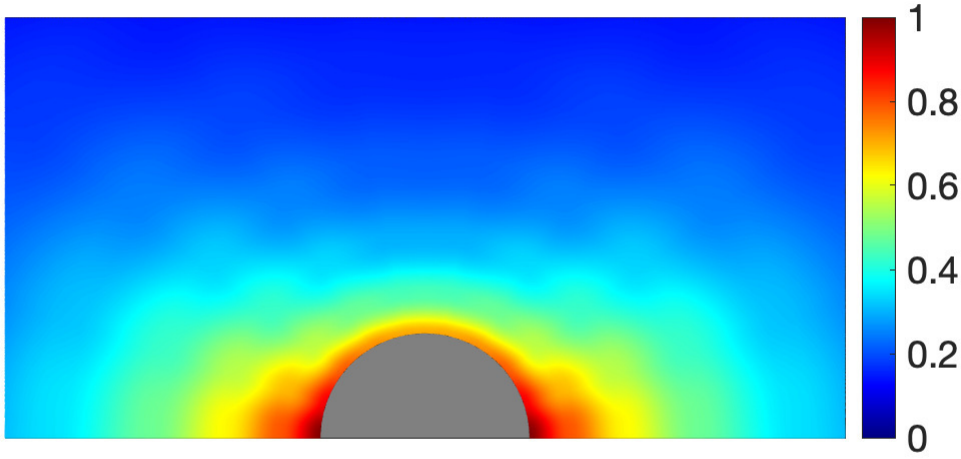}&
\includegraphics[width=2in, scale=0.8]{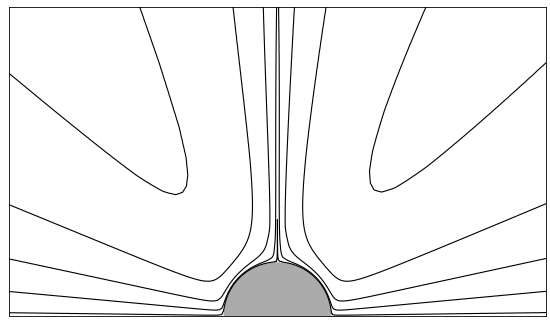}\\[5pt]
(c)&(d)
\end{tabular}
\end{center}
\caption{(Colour online) Steady-state concentration profile and streamlines of the flow around the autophoretic particle at $\Pen=20$ [(a),(b)] and (b) $\Pen=30$ [(c),(d)]. The flow and concentration fields are symmetric about the horizontal axis; therefore only half the particle is shown. The colour bar indicates the value of the solute concentration $c$.}
\label{fig:conc_sfn_map}
\end{figure}

We analyze the dynamics of the autophoretic particle as a function of the P\'{e}clet number.~\citet{Michelin2013} have shown that the particle remains stationary until a critical P\'{e}clet number, $\Pen_{\text{c}}=4$. This range, $0\leq\Pen\leq4$, may be termed as the quiescent regime. As the P\'{e}clet number is increased beyond 4, the fore-aft symmetry in the concentration profile is broken, and the particle executes steady rectilinear self-propulsion. In fig.~\ref{fig:steady_swim}, the steady self-propulsion speed $U$ from our computations is plotted as a function of \Pen, and it is seen that $U$ attains a maximum at $\Pen\approx9$, and decreases smoothly up to a value of $\Pen\approx26$. This range, $4<\Pen\leq26$, is classified as the steady self-propulsion regime. The good agreement of our results with data from~\citet{Michelin2013} lends confidence to the numerical procedure used in the present work. 

In the inset to fig.~\ref{fig:steady_swim}, the self-propulsion speed in the vicinity of symmetry breaking is plotted as a function of $\Pen$, and is found to vary linearly as $U=\left(\Pen-4\right)/16$, in agreement with asymptotic predictions~\citep{Morozov2019,Saha2021} of this singular pitchfork bifurcation. While we have used a value of $R_{o}=100$ for the majority of our simulations, we found that a higher value, $R_{o}=1000$, is required near the bifurcation threshold, in order to correctly model the remote outer region (at distances of order $\left(\Pen-4\right)^{-1}$) in which advection plays a comparable role to diffusion~\citep{Schnitzer2022}. As discussed by~\citet{Schnitzer2022}, factors such as a finite-size domain and bulk-reaction resulting in the consumption of solute (see also~\citet{Farutin2021}) could regularize the bifurcation, leading to a square-root scaling $U\sim\left(\Pen-\Pen_{\text{c}}\right)^{1/2}$ as observed by~\citet{Li2022} and~\citet{Hu2022}. In particular, ~\citet{Li2022} uses an \textit{ad hoc} approximation of the base state around a circular autophoretic drop, where the diffusive concentration profile is set to zero at a finite distance $R_{o}$, to circumvent the issue that there is no steady unbounded solution to Laplace's equation in two dimensions. This represents a finite-size regularization of the pitchfork bifurcation, because the base state concentration profile is always unsteady in the two-dimensional case: at small $\Pen$ there is an outer region at distances of order $1/\Pen^{1/2}$ in which unsteadiness balances diffusion (see e.g.~\citet{Yariv2020}). We believe this is the reason for the square root scaling observed in ~\citet{Li2022}. In three dimensions one has to choose $R_{o}$ to be larger than $1/(\Pen-4)$ so that the advective outer region is included; we suggest that~\citet{Hu2022} did not do this, and this is why they also found a square root scaling. We view our computations being in agreement with the predictions of~\cite{Morozov2019} and~\cite{Saha2021} as further evidence for their accuracy. 

\begin{figure}
  \centerline{\includegraphics[width=5in,height=!]{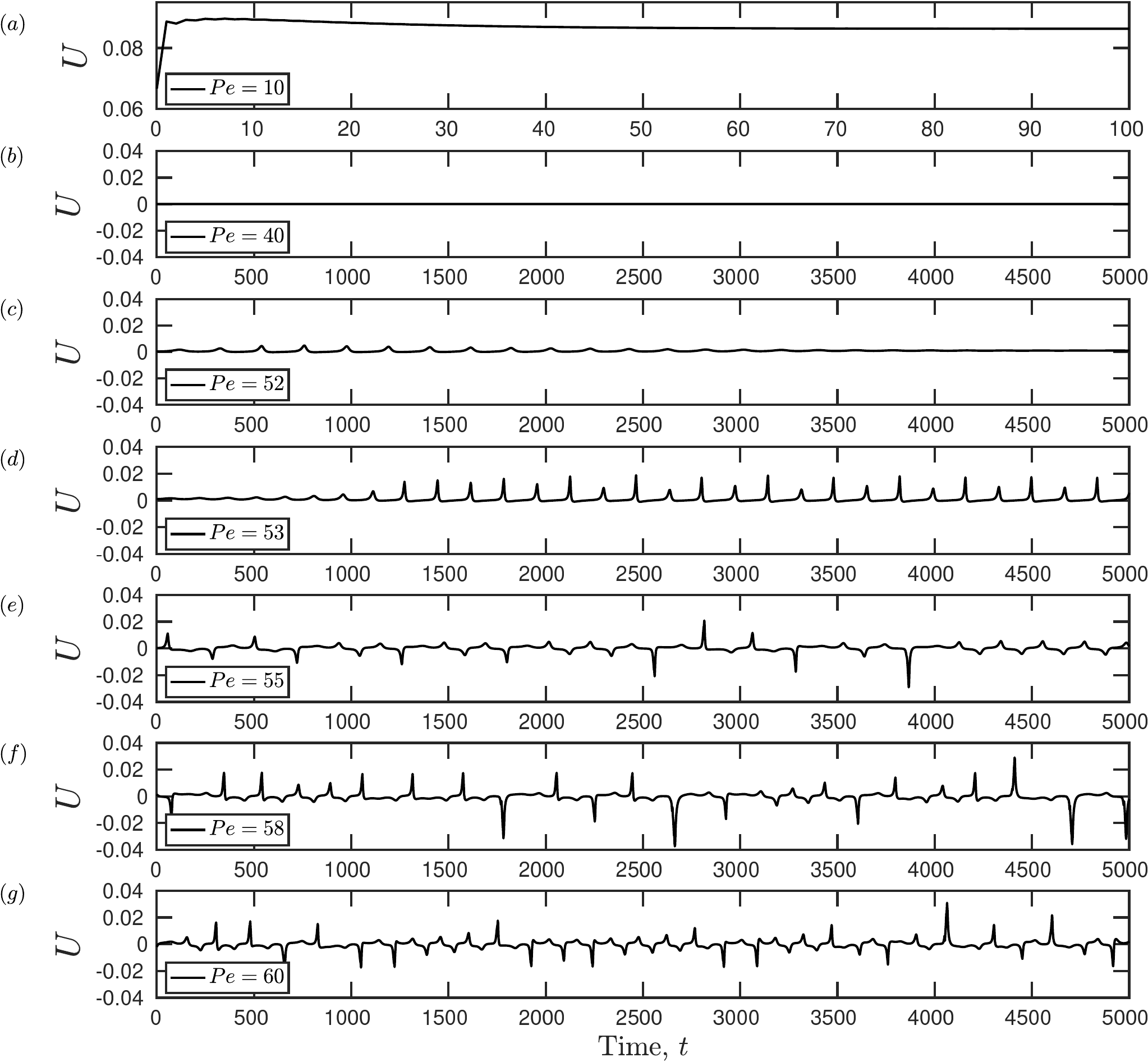}}
  \caption{ (Colour online) Time evolution of the self-propulsion speed at (a) $\Pen=10$, (b) $\Pen=40$, (c) $\Pen=52$, (d) $\Pen=53$, (e) $\Pen=55$, (f) $\Pen=58$, (g) $\Pen=60$.}
\label{fig:u_vs_time}
\end{figure}

At $Pe\approx27$, there is a qualitative change in the particle motion, as it experiences a drop of almost three orders of magnitude in its self-propulsion speed, becoming practically stationary and entering what we refer to as a stirring regime. The seemingly abrupt transition in particle motion may be understood by considering the relative growth rates of the various angular modes near the P\'{e}clet number at which the transition is observed. In fig.~1 of ~\citet{Michelin2013}, analytical solutions to the growth rates of the various unstable modes are plotted as a function of the P\'{e}clet number. Near $Pe\approx30$, the growth rates of the higher order modes are seen to outweigh the dipolar self-propulsion mode, and this could be a reason for the qualitative change in the dynamics of the particle from the swimming to the stirring regime. The transition may also be examined by a comparison of the concentration profile of the solute cloud, and the streamlines of the flow profile, at two different values of $\Pen$, as illustrated in fig.~\ref{fig:conc_sfn_map}. The solute distribution around the autophoretic particle is fore-aft asymmetric at $\Pen=20$, but is nearly fore-aft symmetric at $\Pen=30$ with a maximum concentration at the front and rear stagnation points.  Furthermore, the streamlines in fig.~\ref{fig:conc_sfn_map}~(d) indicate a recirculation of the fluid around the autophoretic particle, where fluid is brought in along the polar axis and expelled at the equator. This flow pattern is consistent with the dominance of a quadrupolar disturbance to concentration profile. This region of parameter space $27\leq \Pen\leq 50$ is therefore analogous to the symmetric extensile pumping regime identified by~\cite{Morozov2019}, wherein the activity of the particle results in the stirring of the fluid around it, without resulting in its self-propulsion. 

\begin{figure}
  \centerline{\includegraphics[width=3.5in,height=!]{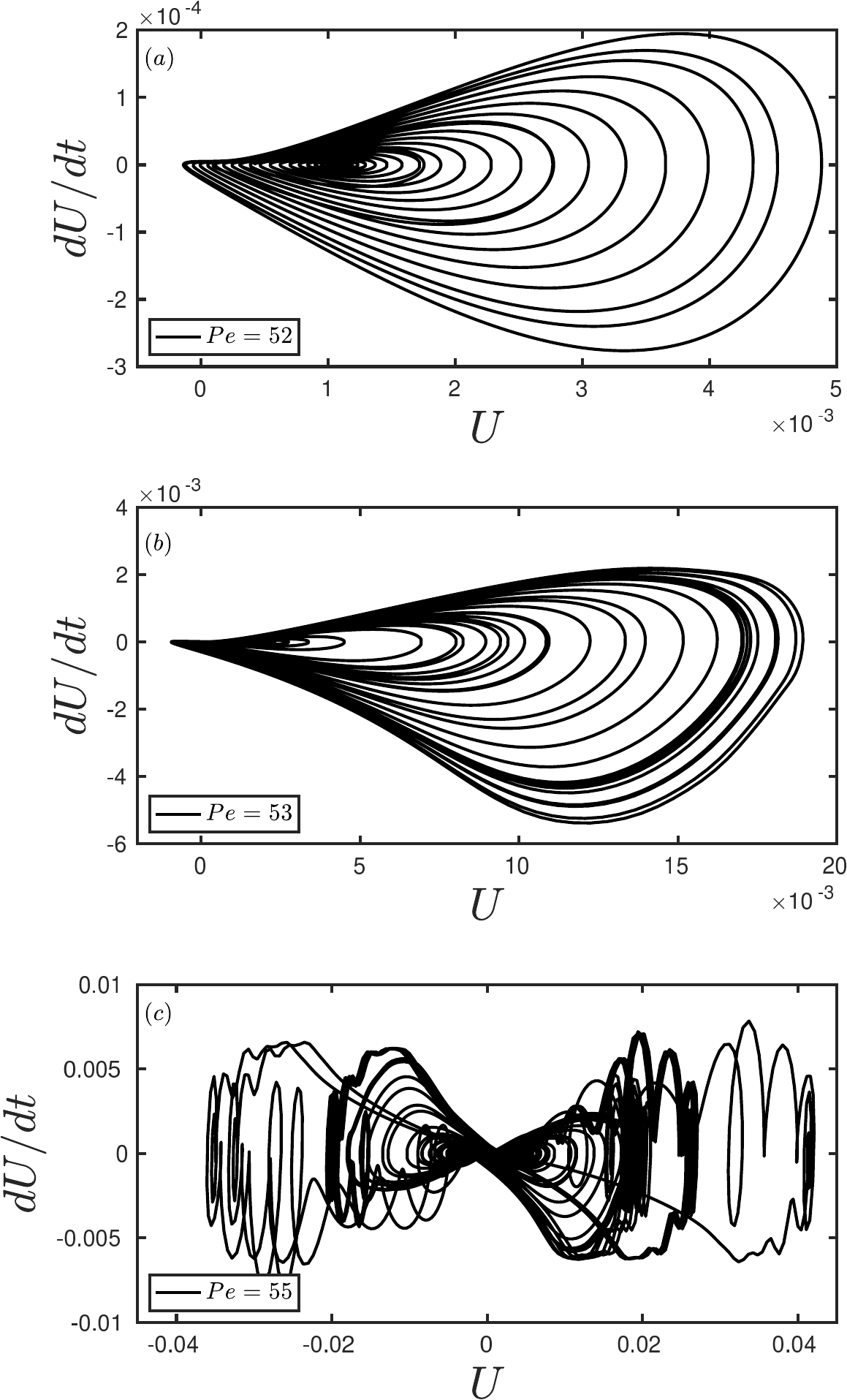}}
  \caption{ (Colour online) Phase-plane diagram at (a) $\Pen=52$, (b) $\Pen=53$, and (c) $\Pen=55$.}
\label{fig:phase_plot}
\end{figure}

In figure~\ref{fig:u_vs_time}, the instantaneous self-propulsion speed of the autophoretic particle is plotted as a function of time for various values of $\Pen$. The swimming speed in the steady regime, as seen from fig.~\ref{fig:u_vs_time}~(a), settles to a constant value of $\textit{O}(10^{-1})$ following an initial transient, while that in the stirring regime (shown in fig.~\ref{fig:u_vs_time}~(b)) is orders of magnitude lower. Beyond $\Pen=51$, an onset of oscillations in the swimming speed is observed. While these oscillations are transient and vanish at long times for $\Pen=52$, they become persistent at $\Pen=53$, as seen from figures~\ref{fig:u_vs_time}~(c) and (d), respectively. With a further increase in $\Pen$, the particle begins to move back and forth along the $z$-axis, the time scale for the reversal in self-propulsion direction decreases and the magnitude of the instantaneous self-propulsion speed increases, as seen from figures~\ref{fig:u_vs_time}~(e)-(g). The region $\Pen\geq55$ may be termed as the chaotic regime, characterized by short bursts of self-propulsion in arbitrary directions (along the $\pm z$-axis) and sharp changes in both the magnitude and direction of the self-propulsion speed. {This transition to chaos occurs through a period doubling cascade, as illustrated through the phase-plane diagram $\left(dU/dt,U\right)$ in figure~\ref{fig:phase_plot}. The pattern of closed loops in figures~\ref{fig:phase_plot}~(a) and (b), corresponding to $\Pen=52$ and $\Pen=53$, respectively, is qualitatively similar to that reported by~\cite{Hu2022}, who also observe a period-doubling route to chaos. As the P\'{e}clet number is increased to $\Pen=55$ [fig.\ref{fig:phase_plot}~(c)] the trajectory resembles the familiar butterfly-shaped pattern observed in chaotic systems~\citep{Strogatz2015}.}

\begin{figure}
\begin{center}
\begin{tabular}{c}
\includegraphics[width=5in, scale=1]{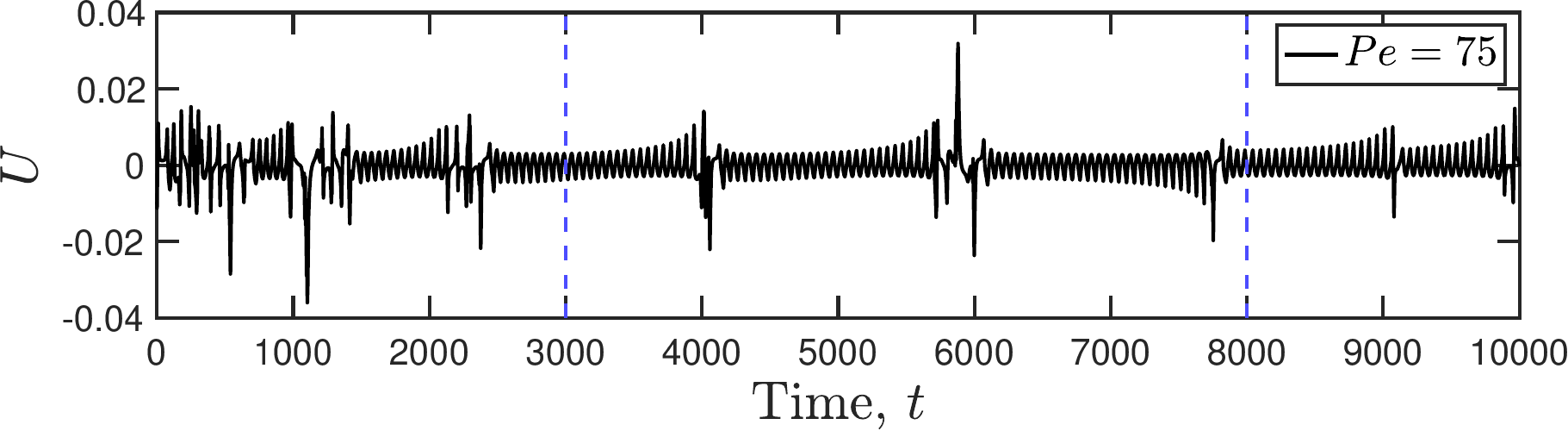}\\
(a)\\
\includegraphics[width=5in, scale=1]{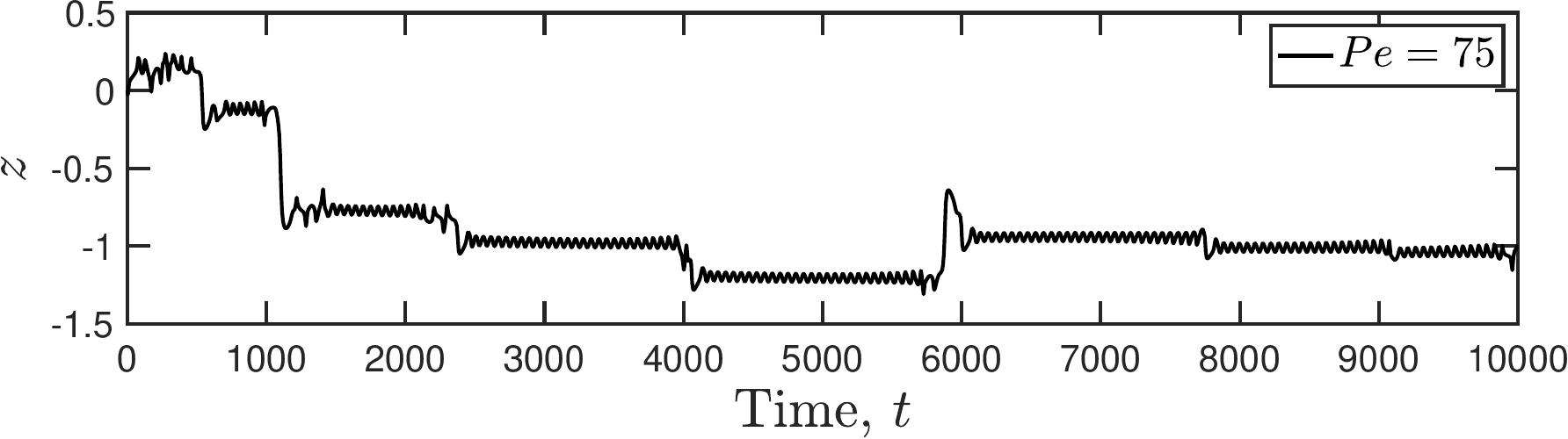}\\[5pt]
(b)\\
\end{tabular}
\begin{tabular}{cc}
\includegraphics[width=2.5in, scale=0.8]{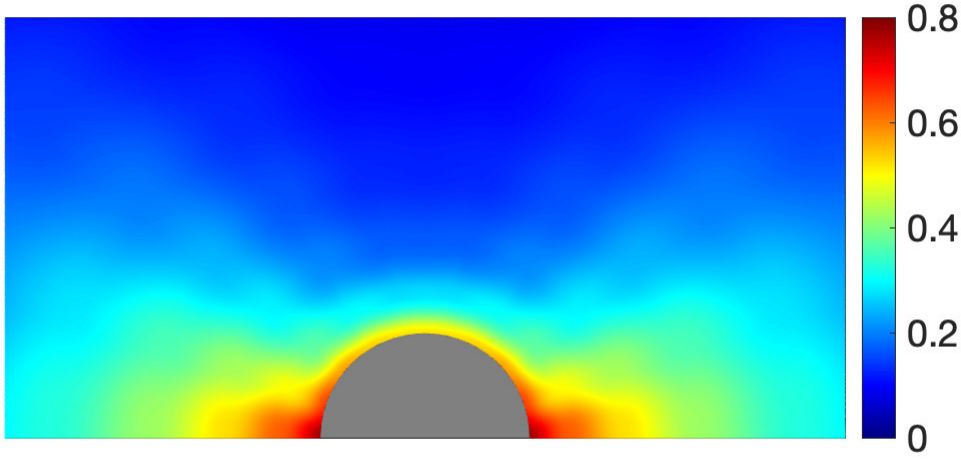}&
\includegraphics[width=2in, scale=0.8]{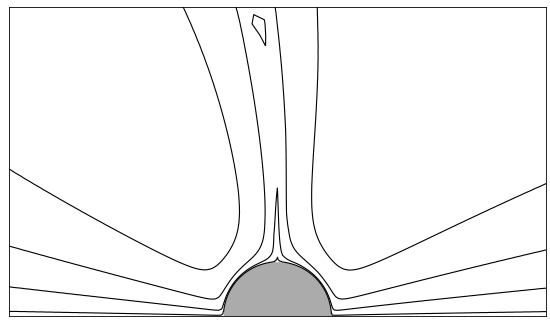}\\[5pt]
(c)&(d)
\end{tabular}
\end{center}
\caption{(Colour online) Time evolution of (a) the self-propulsion speed, and (b) displacement along $z-$ axis, for an autophoretic particle at $\Pen=75$. The complete simulation output in the window $t\in\left(3000,8000\right)$ is saved at intervals of fifty dimensionless time units. The time-averaged concentration profile and streamlines of the flow around the autophoretic particle in this window are shown in (c) and (d), respectively.}
\label{fig:pe75_analysis}
\end{figure}

\begin{figure}
\begin{center}
\begin{tabular}{c}
\includegraphics[width=5in, scale=1]{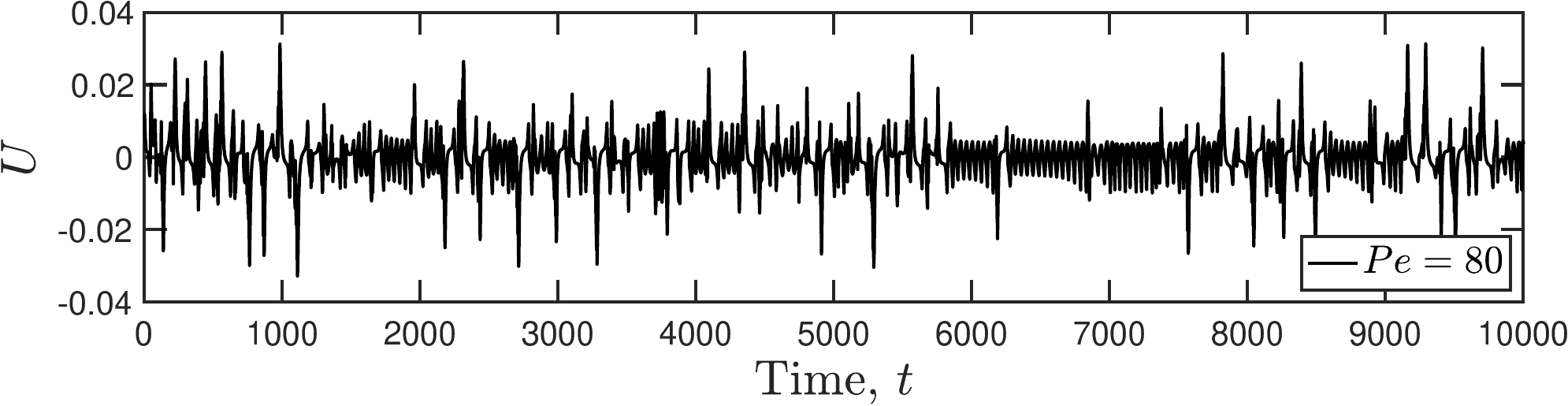}\\
(a)\\
\includegraphics[width=5in, scale=1]{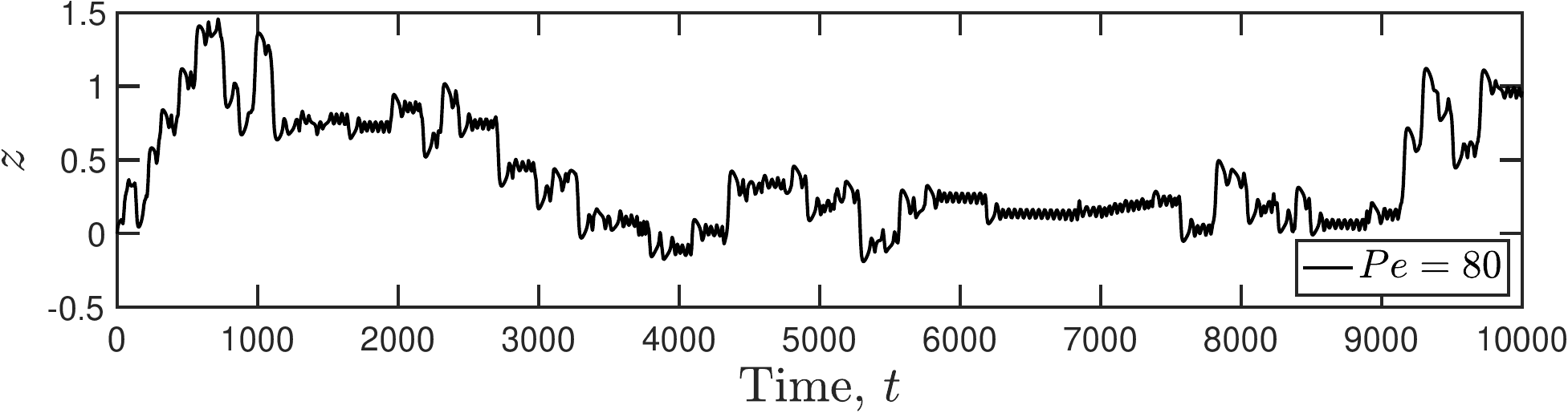}\\
(b)
\end{tabular}
\end{center}
\caption{(Colour online) Time evolution of (a) the self-propulsion speed, and (b) displacement along $z-$ axis, for an autophoretic particle at $\Pen=80$.}
\label{fig:pe80_analysis}
\end{figure}

The dynamics in the chaotic regime is explored further by considering a representative case of $\Pen=75$. The time series of instantaneous self-propulsion speed at $\Pen=75$ is shown in figure~\ref{fig:pe75_analysis}~(a). Following an initial transient that lasts till $t\approx3000$, the velocity settles into a characteristic pattern, with intermittent bursts of chaos that interrupt nearly regular oscillations of slowly varying amplitude~\citep{bpv1984}. The instantaneous position of the particle along the $z-$axis, evaluated using eq.~(\ref{eq:z_pos}), is plotted in fig~\ref{fig:pe75_analysis}~(b) as a function of time. After a transient period, the particle is seen to oscillate about a mean position of $z=-1$. The time-averaged concentration profile and streamlines of flow around the autophoretic particle, evaluated in the window $t\in\left(3000,8000\right)$, are plotted in figs.~\ref{fig:pe75_analysis}~(c) and (d), respectively. The concentration map is largely fore-aft symmetric, and this is consistent with the limited (net) mobility of the particle seen in (b). The pattern of the streamlines is qualitatively similar to that observed in the stirring regime (\textit{cf}. fig.~\ref{fig:conc_sfn_map}~(d)). A similar behaviour is observed at $\Pen=80$, following an initial transient period that lasts till $t\approx1500$. The intermittent bursts of chaos are more frequent, however, as observed in figure~\ref{fig:pe80_analysis}~(a). Due to the more frequent injections of this `turbulent' motion, the particle travels a larger distance than at $\Pen=75$, as seen from figure~\ref{fig:pe80_analysis}~(b).  While the velocity and displacement time series of the autophoretic particle in the chaotic regime vary markedly depending upon the P\'{e}clet number, an analysis of the particle's mean square displacement and velocity autocorrelation reveals certain unifying features, as discussed below.

\begin{figure}[t]
  \centerline{\includegraphics[width=4.0in,height=!]{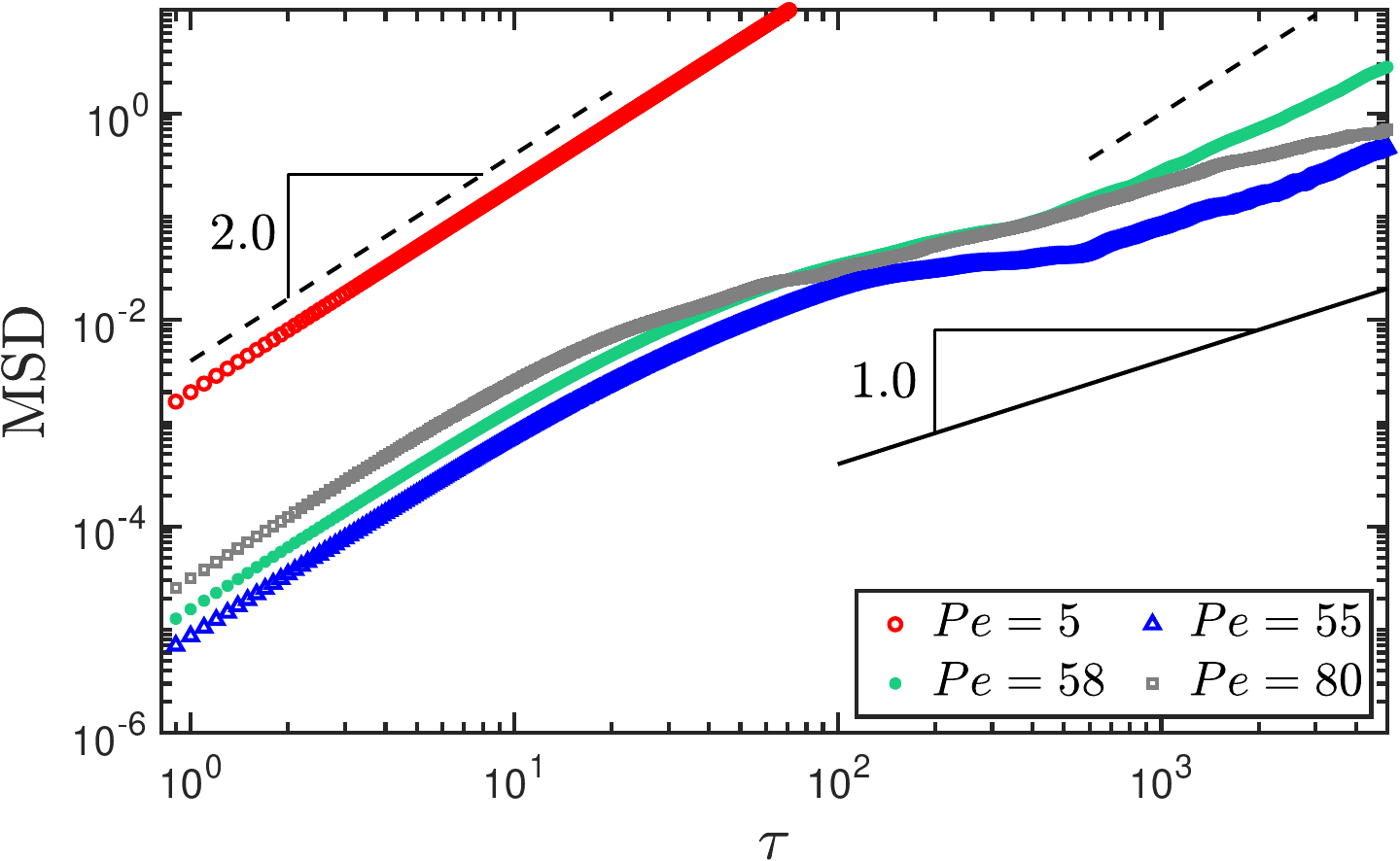}}
  \caption{ (Colour online) Mean square displacement of an autophoretic particle at various P\'{e}clet numbers. Broken lines indicate a slope of $2$, while solid line has a slope of $1$.}
\label{fig:msd_mult}
\end{figure}

\begin{figure}[t]
\begin{center}
\begin{tabular}{cc}
\includegraphics[width=0.45\linewidth]{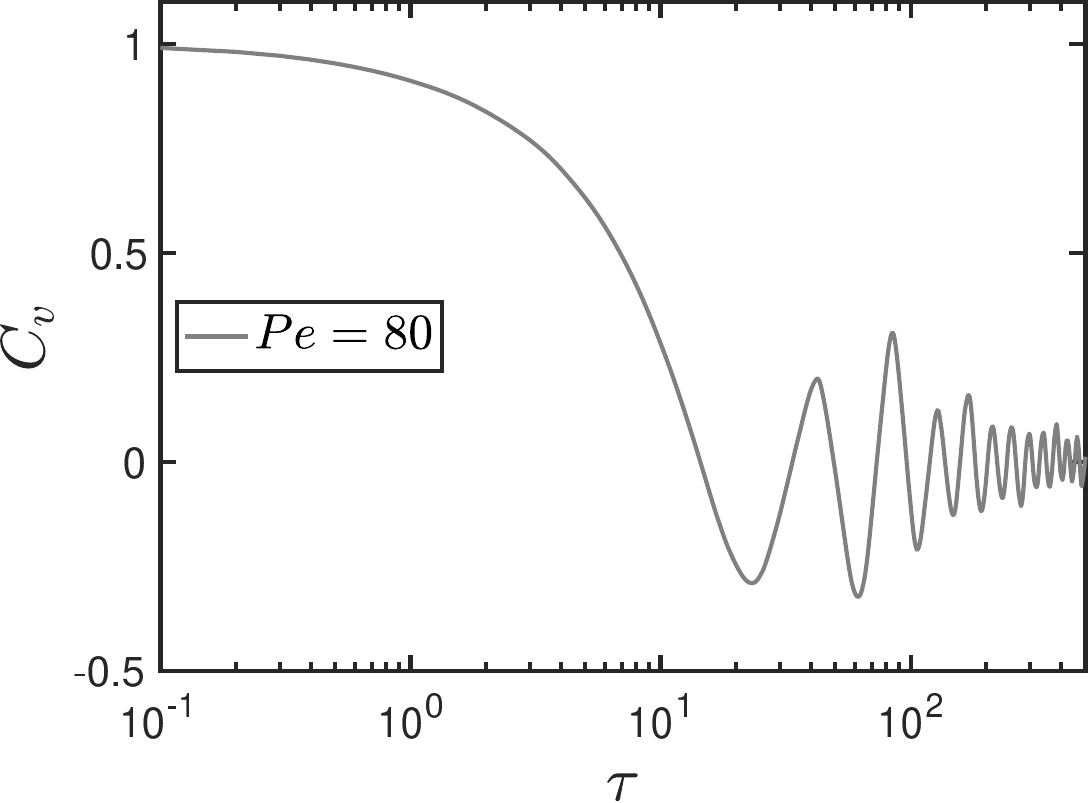}&
\includegraphics[width=0.45\linewidth]{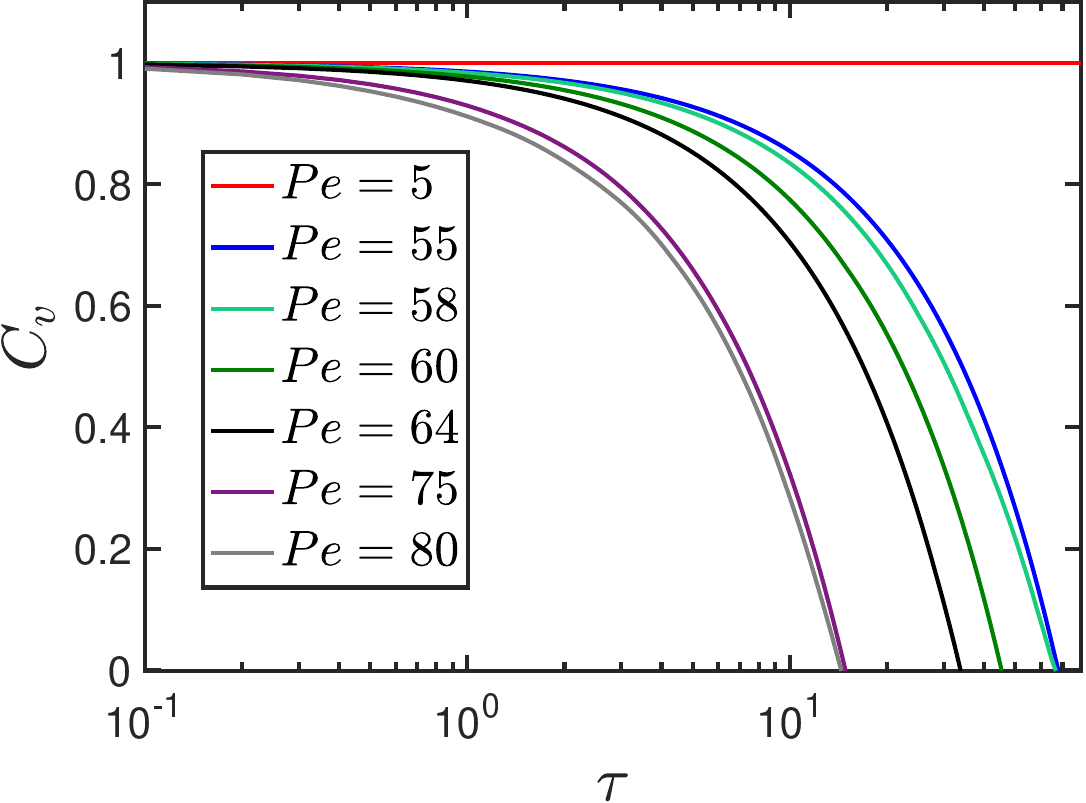}\\
(a) & (b)
\end{tabular}
\end{center}
\caption{(Colour online) Velocity autocorrelation of an active particle (a) at $\Pen=80$, and (b) over a range of P\'{e}clet numbers.}
\label{fig:vac_comparison}
\end{figure}

In figure~\ref{fig:msd_mult}, the MSD of the particle is plotted as a function of the lag time for a range of P\'{e}clet numbers.  At $\Pen=5$, corresponding to steady self-propulsion, the MSD grows as $\sim \tau^2$, since the particle moves rectilinearly at a constant speed.  P\'{e}clet numbers in the range $27\leq\Pen<50$ are not examined since the particle is practically stationary, with only the fluid around it undergoing symmetric pumping, or stirring. At $\Pen=55$ and $\Pen=58$, the mean square displacement grows quadratically with the lag time at early times, followed by a region of slower growth, before appearing to tend toward a long-time ballistic regime. This evolution may be qualitatively compared to the findings by~\cite{Hu2022}, who observed for $Pe=54$ that the MSD has an early-time and late-time ballistic regime. At $\Pen=80$, however, only an early-time ballistic motion is observed, followed by a transition to what more closely resembles a diffusive regime (MSD $\sim \tau$). While such a trend is not observed in axisymmetric simulations of~\cite{Hu2022}, who do not report computational results for $\Pen>60$, it is similar to the transition to diffusive motion observed their 3D simulations, albeit at lower values of the P\'{e}clet number ($\Pen\approx24.5$). The different MSD behaviors observed for $\Pen=55$ and $\Pen=80$ can be interpreted in terms of the differences in the velocity time series at these two P\'{e}clet numbers: the time series at $\Pen=55$ (fig.~\ref{fig:u_vs_time}~(e)) is within the window of period doubling identified by~\cite{Hu2022}, whereas that for $\Pen=80$ contains bursts of intermittent chaos (figure~\ref{fig:pe80_analysis}~(a)). Thus, we conjecture that the long-time diffusive motion of an autophoretic particle, whether constrained to axisymmetric motion or free to move in three dimensions, is due to intermittency in its chaotic dynamics.

In fig~\ref{fig:vac_comparison}~(a), the VAC function for an autophoretic particle with $\Pen=80$ is plotted as a function of time. Rapid changes in the direction of self-propulsion result in a decorrelation of the signal, which crosses zero at $\tau\approx15$. This is followed by the appearance of negative dips and oscillations in the VAC: these features have also been observed experimentally by~\cite{Suda2021} and~\cite{Hokmabad2021}, as discussed below in detail. The variation in the VAC with $\Pen$ is illustrated in fig~\ref{fig:vac_comparison}~(b), where the data for $\Pen\geq55$ have been shown only till the first instance of their crossing zero, for clarity. A decrease in the 
P\'{e}clet number from $\Pen=80$ results in less rapid changes in the particle motion, which is reflected in the longer time required for the velocity to decorrelate. Finally, at $\Pen=5$, the VAC is constant in time, indicating a complete correlation in the velocity, due to the persistent unidirectional translation of the particle. Such a decrease in the correlation time with an increase in $\Pen$ has also been reported by~\cite{Chen2021}, who performed 3D simulations of a phoretic particle using the immersed boundary method. That study, however, does not report a zero crossing in the VAC for the range of lag times investigated. 

Experiments on water droplets in a mixture of oil and surfactant~\citep{Suda2021}, as well as oil droplets in an aqueous surfactant solution~\citep{Hokmabad2021}, have both shown that the decorrelation in the droplet velocity is faster at larger values of $\Pen$, in qualitative agreement with the trends in fig~\ref{fig:vac_comparison}~(b). In these experiments,  the droplets undergo self-propulsion due to a Marangoni stress imbalance at the interface, driven by a reaction between the droplet contents and the surfactant, resulting in the generation of oil-filled micelles. The active droplets undergo a transition from ballistic motion to anomalous diffusion with an increase in the P\'{e}clet number~\citep{Hokmabad2021}, which is accompanied by a faster decay in the VAC. The droplets are found to avoid the chemical trail that they generate, and undergo a two-dimensional self avoiding walk, with MSD $\sim\tau^{3/2}$. This route for trail avoidance is not possible for the axisymmetric model considered here, and is a major point of distinction from the two-dimensional experiments discussed above. This could also potentially be the reason for the slower growth in the MSD at long times, leading to a near-diffusive regime observed at $\Pen=80$.

\section {Forced autophoretic particle}
\label{sec:ep_fin_results}

\begin{figure}[t]
\begin{center}
\begin{tabular}{cc}
\includegraphics[width=0.45\linewidth]{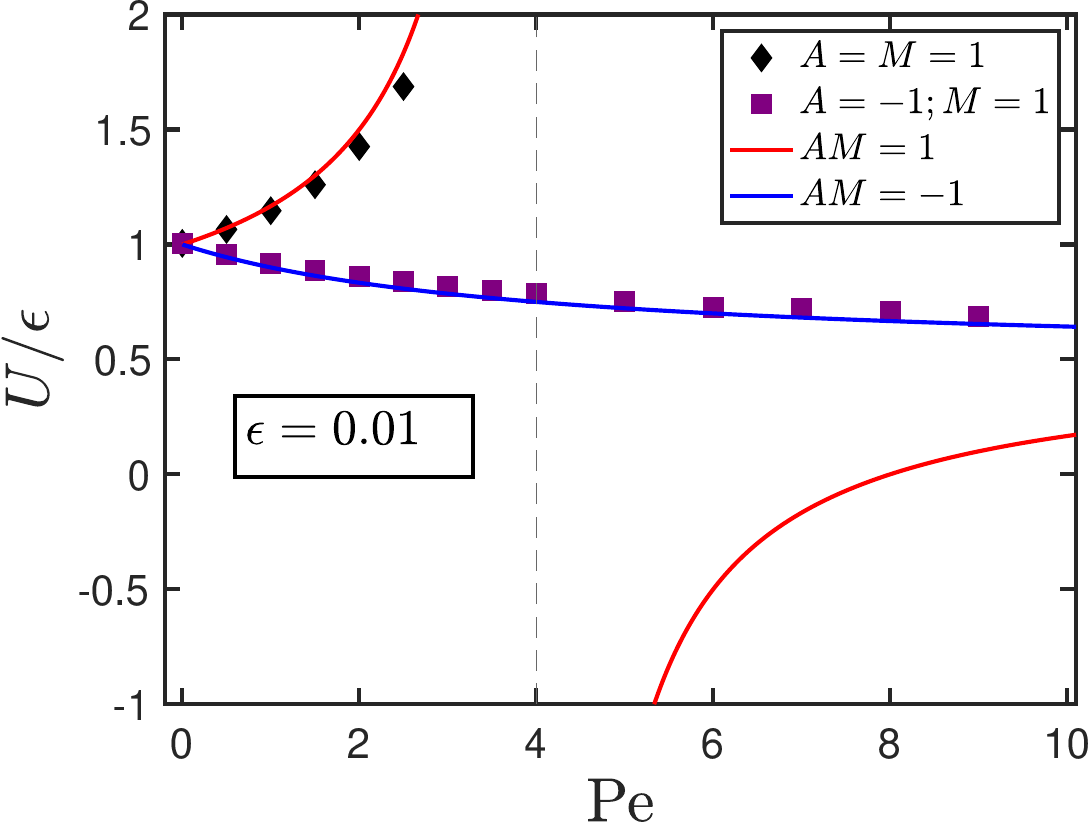}&
\includegraphics[width=0.45\linewidth]{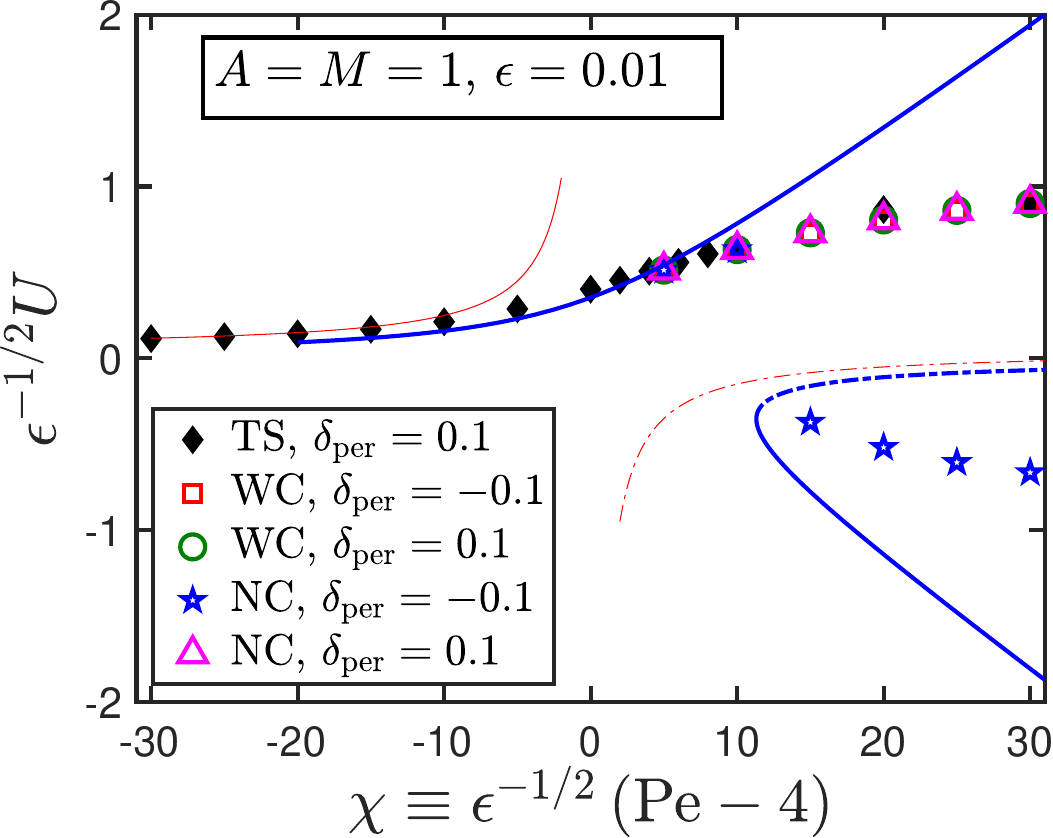}\\
(a) & (b)
\end{tabular}
\end{center}
\caption{(Colour online) Scaled self-propulsion speed of a forced autophoretic particle with $\epsilon=0.01$. Symbols are numerical results, and in (a) lines are analytical predictions from the global analysis of~\citet{Saha2021}; in (b) thick lines denote solutions from their local analysis, and thin lines represent the global solution. Solid lines denote the stable solution branch, while the broken lines represent the unstable branch. Filled symbols indicate solutions obtained using the transient solver (TS), while hollow symbols denote results obtained using the iterative solver. Results obtained with and without a continuation scheme have been denoted by the abbreviations ``WC'' and ``NC'' respectively.}
\label{fig:comp_schnitz}
\end{figure}

~\citet{Saha2021} have derived asymptotic approximations to the motion of an autophoretic particle subjected to a weak external force ($\epsilon\ll1$). Specifically, they have obtained `global' and a `local' approximations to this problem. 
In the global approximation an $O(\epsilon)$ perturbation is constructed to the trivial solution of an unforced particle, where the solute concentration is isotropic and there is no flow.
This leads to the prediction $U\sim \epsilon(8-AM\Pen)/[2(4-AM\Pen)]$ as $\epsilon\to 0$.
Thus, increasing $\Pen$ for $AM=-1$, where the trivial solution is stable, leads to a reduction in the speed of particle translation due to diffusio-osmotic flow around the particle.
The case of $AM=1$ is more interesting: here the instability of the unforced problem at $\Pen=4$ is manifested as a divergence in the global approximation at this value of $\Pen$.
A comparison of their global prediction against our numerical solution is shown in figure~\ref{fig:comp_schnitz}~(a). 
For $AM=1$ the numerical result matches well with the approximate global solution at small values of $\Pen$.
For $AM=-1$ the global approximation and the numerical results agree over a larger range of the P\'{e}clet number. 

~\citet{Saha2021} have also derived a local approximation for the particle translation speed, valid in the vicinity of the bifurcation point of the unforced problem, $\Pen=4$. 
More specifically, this approximation is valid when the variable $\chi\equiv\epsilon^{-1/2}\left(\mathrm{Pe}-4\right)=O(1)$ as $\Pen\to 4$; that is, $\chi$ represents the appropriate deviation from the bifurcation point. 
The results for the particle speed are given in equations (8.1) to (8.3) of~\citet{Saha2021} and not repeated here for brevity.
In summary, the speed is of $O(\epsilon^{-1/2})$ near the bifurcation point, and there are two stable solutions branches: one corresponds to $U>0$ and smoothly matches to the global approximation as $\chi\to\-\infty$; and the other corresponds to $U<0$ and only exists for $\chi>8\sqrt{2}$. 
There is also an unstable branch with $U<0$ for $\chi>8\sqrt{2}$.
This is the classic scenario of an imperfect (singular) pitchfork bifrucation.
The stable branch with $U<0$ suggests that steady propulsion is possible in a direction anti-parallel to the external force, which is remarkable.
On this note, experiments on the sedimentation of active droplets in a semicylindrical glass container~\citep{Moerman2019} show that, above a critical P\'{e}clet number, the droplets move \textit{transiently} against the direction of the biasing force (i.e., gravity). 
We are not aware, however, of experiments in which an active particle/drop undergoes \textit{steady} self-propulsion against the direction of the external force.

In figure~\ref{fig:comp_schnitz}~(b), the particle speed for $AM=1$ computed from the present work, scaled by $\epsilon^{-1/2}$, is compared against their local approximation. 
The agreement is good near the bifurcation point ($\chi=0$). 
Four different numerical scenarios are examined in figure~\ref{fig:comp_schnitz}~(b) for the iterative solver corresponding to various choices of $\delta_{\text{per}}$ and whether or not a continuation scheme is applied, wherein the converged concentration field at a particular value of the P\'{e}clet number is used as the initial guess for the computation at the next higher value of $\Pen$. 
At $\delta_{\text{per}}=0.1$, the numerical predictions of the iterative solver are identical, irrespective of whether continuation is employed. 
With $\delta_{\text{per}}=-0.1$, contrasting trends are observed: using a continuation scheme, the predictions agree with those obtained for the $\delta_{\text{per}}=0.1$ case. 
However, in the absence of continuation, a stable steady steady self-propulsion is observed in the direction opposite to the external forcing, for $\chi\geq15$. 
Using the converged concentration field at these P\'{e}clet numbers as the initial condition for the transient solver, however, predicts a steady phoretic speed in the direction of the external forcing, i.e. a jump from the anti-parallel to parallel branches of the local approximation.  
Additionally, the results predicted by the transient solver for the range of $\Pen$ examined in figure~\ref{fig:comp_schnitz}~(b) are unaltered by the choice of $\delta_{\text{per}}$, or the use of a continuation scheme. 
These results only lead to motion with $U>0$.

\begin{figure}[t]
  \centerline{\includegraphics[width=4.0in,height=!]{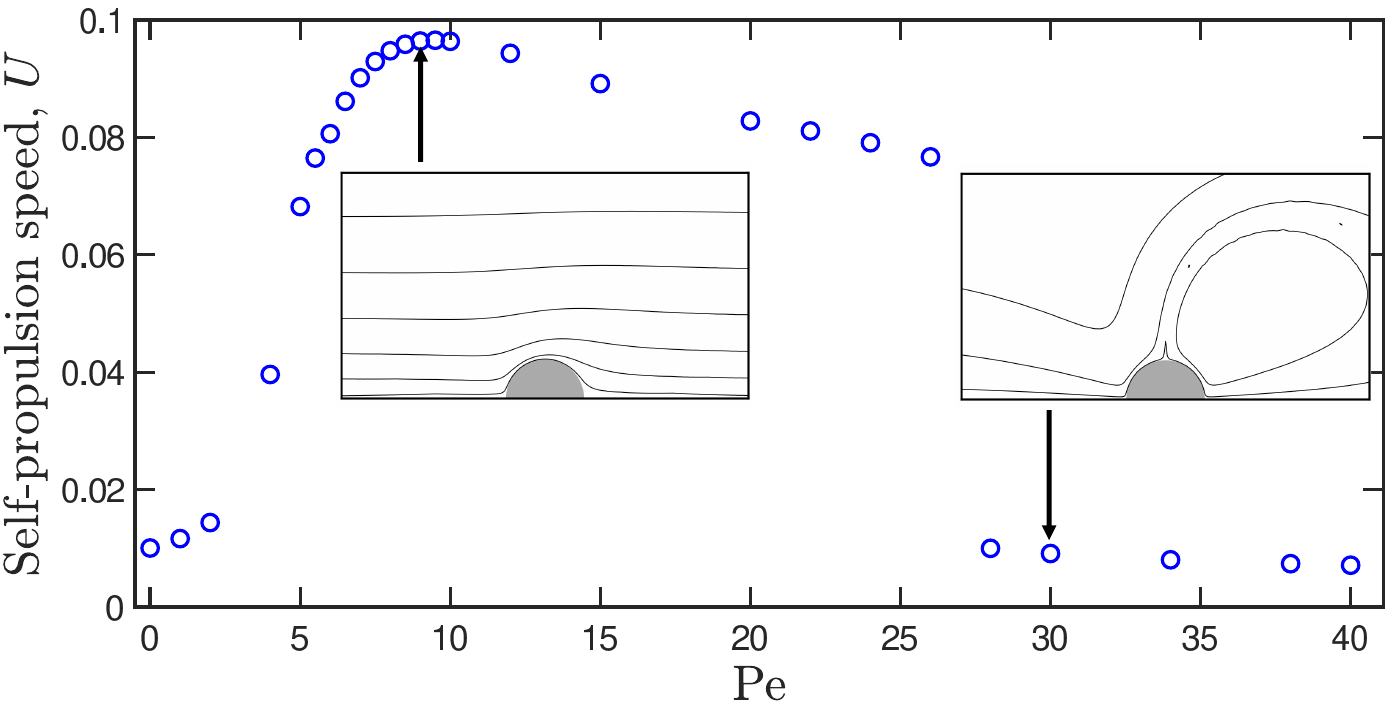}}
  \caption{ (Colour online) Translation speed of an autophoretic particle with $A=M=1$ and $\epsilon=0.01$. The insets show the streamlines of the flow around the particle at $\Pen=9$ and $\Pen=30$.}
\label{fig:finite_eps_speed_u}
\end{figure}

\begin{figure}
  \centerline{\includegraphics[width=5in,height=!]{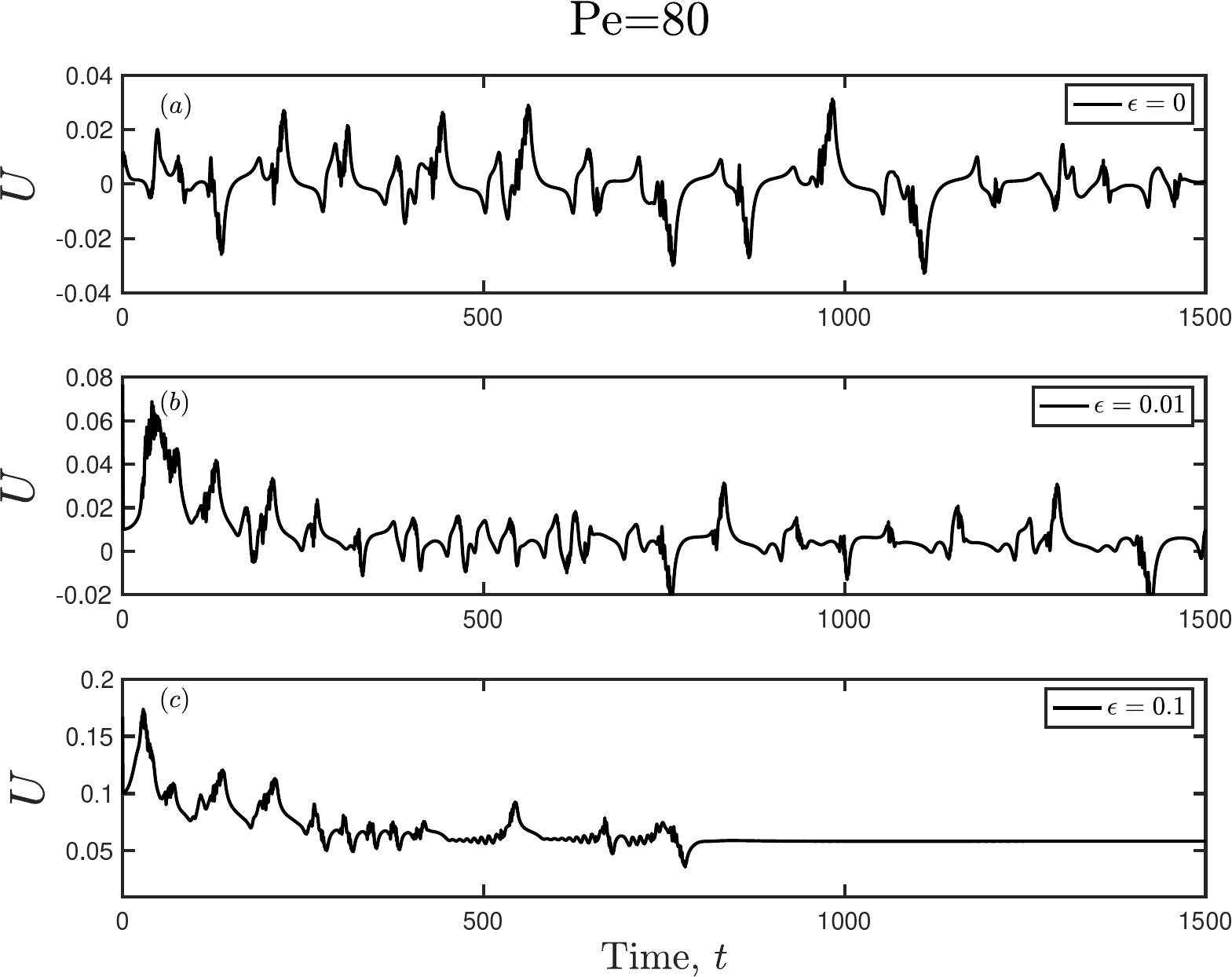}}
  \caption{ (Colour online) Transient translation speed of an autophoretic particle with $\Pen=80$ and (a) $\epsilon=0.0$, (b) $\epsilon=0.01$, and (c) $\epsilon=0.1$.}
\label{fig:stead_ext_force}
\end{figure}

\begin{figure}
  \centerline{\includegraphics[width=5in,height=!]{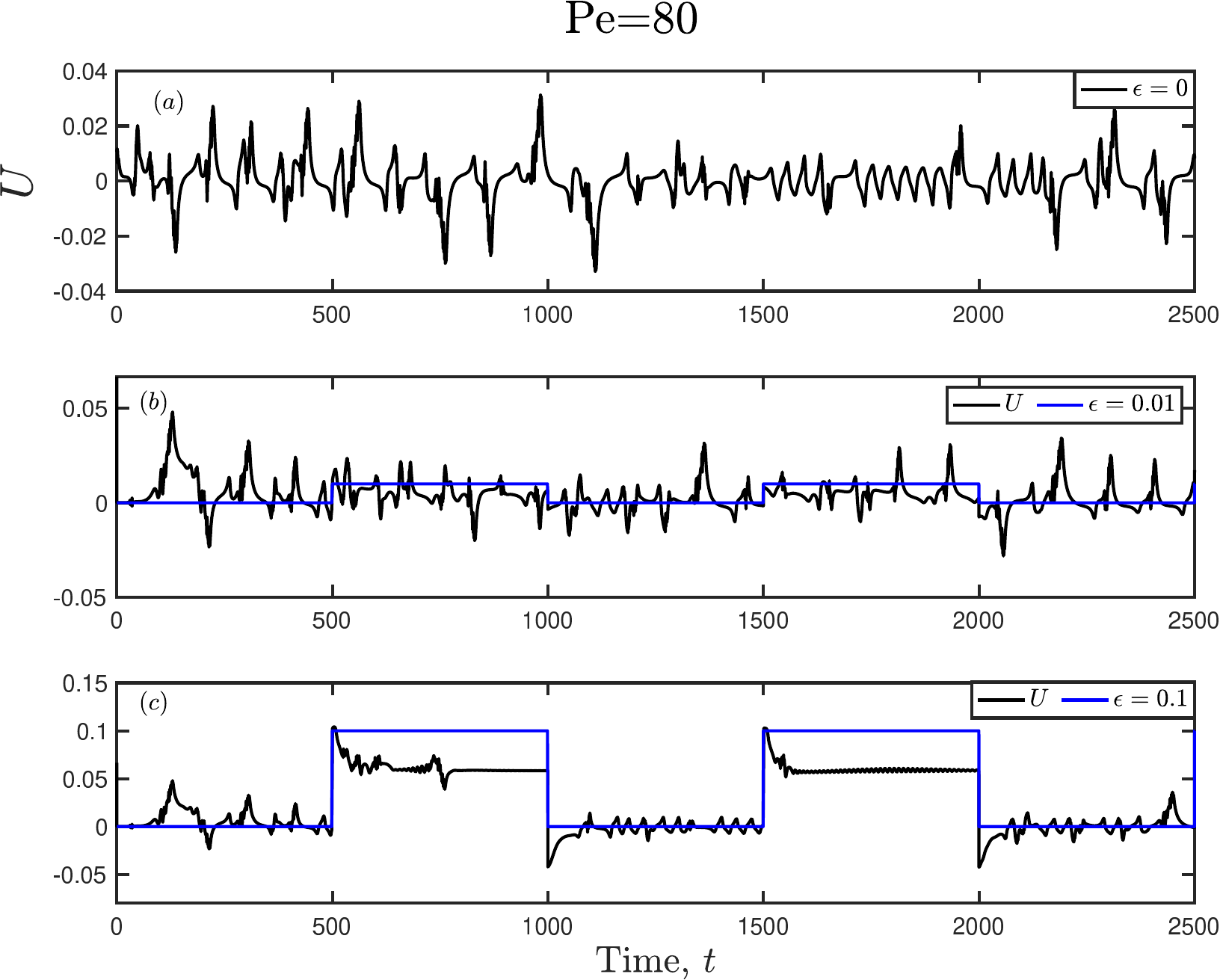}}
  \caption{ (Colour online) Transient translation speed of an autophoretic particle with $\Pen=80$ and (a) $\epsilon=0.0$, (b) $\epsilon=0.01$, and (c) $\epsilon=0.1$. (a) corresponds to an unforced particle, while the time-dependent forcing has been indicated by blue lines in (b) and (c).}
\label{fig:tdep_ext_force}
\end{figure}

The variation of the steady self-propulsion speed for a particle with $AM=1$ and $\epsilon=0.01$ over a larger  range of P\'{e}clet number is shown in figure~\ref{fig:finite_eps_speed_u}. There is no quiescent regime for such a particle, and the self-propulsion speed increases with the P\'{e}clet number upto $\Pen\approx10$, before decreasing. 
At $\Pen\approx28$, an abrupt decrease in the self-propulsion speed is observed, which is accompanied by a recirculation region in front of the particle. This represents a perturbed form of the stirring regime observed for an unforced particle, where the fore-aft symmetry in the flow profile is due to the weak external force.
The qualitative similarity of the particle speed variation with $\Pen$ to that of an unforced particle suggests that chaotic dynamics should occur at sufficiently large $\Pen$, and this is indeed the case.
In figure~\ref{fig:stead_ext_force}, the effect of a steady force on the chaotic dynamics of an autophoretic particle is illustrated, at a representative value of $\Pen=80$ for $\epsilon=0, 0.01$ and $0.1$. 
Increasing the value of $\epsilon$ quenches the chaos, as seen for $\epsilon=0.1$, where the particle attains a steady translation speed after an initial period of transient chaotic self-propulsion. 
However, at $\epsilon=0.01$ the magnitude of the external force is not sufficient to quench the chaos.
Nonetheless, following a transient period that lasts till $t\approx 1000$, the time-averaged value of the self-propulsion speed is observed to increase commensurately, relative to the unforced case, with the magnitude of the external force. 

In figure~\ref{fig:tdep_ext_force}, we demonstrate the effect of a time-dependent external force on the chaotic dynamics. 
The external force is varied between zero and a fixed value of $\epsilon$ over a time interval $t_{\text{box}}=500$. 
The difference in the average self-propulsion speed between adjacent blocks computed over the interval $t_{\text{box}}$ is $\approx\epsilon/2$ for both the non-zero values of the dimensionless external force considered in figure~\ref{fig:tdep_ext_force}.
At $\epsilon=0.1$ the external force is of sufficient strength to eventually quench the chaotic dynamics during the time when the external force switched on.
Thus, even a relatively weak external force can significantly alter the dynamics of an autophoretic particle within the chaotic regime.
Our computations are axisymmetric, which restricts the particle motion to be along the direction of the external force.
However, we expect our conclusions to hold qualitatively for unconstrained (i.e. three-dimensional) motion.
For example, with reference to~\ref{fig:tdep_ext_force}, a time-dependent external forcing could lead to transient alignment of the particle motion with the direction of the external force, whereas unconstrained chaotic motion occurs during times when the external force is switch off. 

\section {Conclusions}
\label{sec:conclusion}

Using a spectral element method to solve for the velocity and concentration fields around a rigid, spherical autophoretic particle in axisymmetric translation, we have examined the motion of the particle under the presence and absence of an external biasing force. For the unforced particle, we have demonstrated that the scaling of the self-propulsion speed in the vicinity of the transition to spontaneous motion scales linearly with $\left(\Pen-4\right)$. The transition to chaotic dynamics proceeds through quiescent, steady, and stirring regimes, as $\Pen$ is increased. The motion of the particle in the chaotic regime is analyzed using the MSD and the VAC. At sufficiently large $\Pen$ (e.g. $\Pen=80$) intermittent chaos is observed in the velocity time series, and the MSD appears to more closely follow a diffusive, rather than ballistic, scaling at long times. The effect of an external force on the particle dynamics in the chaotic $\Pen$ regime was examined; here, we showed that the chaotic motion may be quenched by modulating the magnitude of the external force.

Admittedly, fluid droplets, and not rigid colloids, are the experimentally realizable versions of chemically active autophoretic entities. However, theoretical and numerical studies on rigid autophoretic particles (which are easier to set up computationally) have made fundamental predictions about the dynamics of active particles, many of which have been found to be in qualitative agreement with experiments on droplets. For example, theoretical studies on rigid autophoretic particles~\cite{Hu2019,Hu2022} predict the existence of stationary, steady self-propulsion, and meandering motion followed by chaos, which have also been observed experimentally for droplets. 
Active droplet systems can also be affected by the presence of an external force field, such as gravity, as discussed by~\cite{Moerman2019}. The recent work by~\citet{Saha2021} investigating the dynamics of autophoretic colloids subjected to a biasing force provides a simple model for understanding the effect of gravity on such systems. We therefore contend that, despite the fundamental physico-chemical differences between droplets and rigid particles, analysis of the latter can provide valuable insights on the dynamics of self-propelled autophoretic systems in general.

A natural next step would be the study of active droplets, in which both diffusiophoresis and Marangoni flow contribute to self-propulsion. As mentioned earlier,~\citet{Morozov2019} introduce a tunable parameter $m$ that dictates the relative importance of diffusiophoretic effects to Marangoni flow. 
The findings by~\cite{Morozov2019} suggest that some amount of diffusiophoresis is needed in order to induce the transition to chaos, thereby implying that only steady translation is observed for purely Marangoni propulsion ($m=0$). 
It would be worthwhile to re-examine the chaotic dynamics in such active droplet systems by adapting the numerical-scheme presented in this work.

\noindent\textbf{Acknowledgements.} R.K. thanks Nicholas Chisholm for detailed discussions regarding the numerical solver. 

\noindent\textbf{Funding.} We gratefully acknowledge the support of the Charles E. Kaufmann Foundation of the Pittsburgh Foundation.

\noindent\textbf{Declaration of interests.} The authors report no conflict of interest.

\appendix

\section{Numerical convergence}
\label{sec:app_a}

\begin{figure}
\begin{center}
\begin{tabular}{cc}
\includegraphics[width=2.5in, scale=0.8]{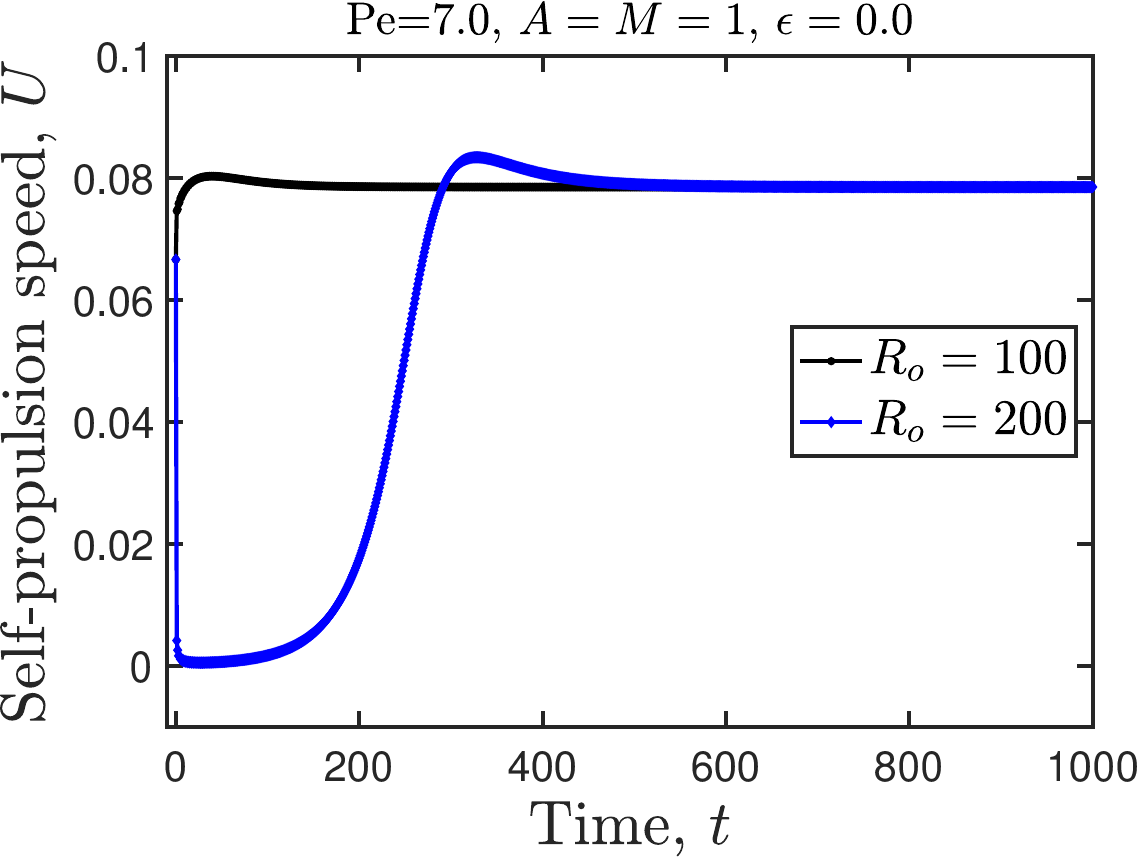}&
\includegraphics[width=2.5in, scale=0.8]{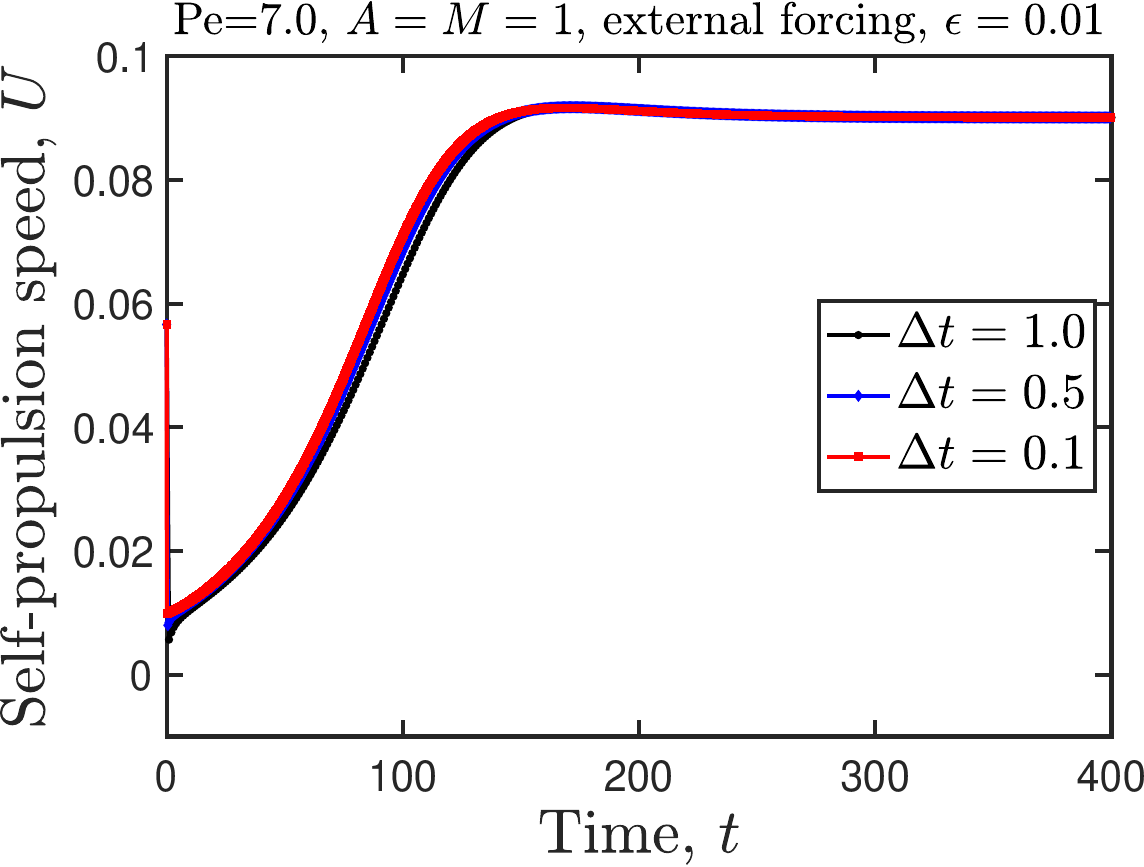}\\
(a) & (b)
\end{tabular}
\end{center}
\caption{(Colour online) Convergence with respect to (a) outer shell radius, at a fixed value of $\Delta t=1.0$ and (b) timestep width, at a fixed value of $R_{o}=100$.}
\label{fig:validn_app}
\end{figure}

The convergence of the numerical results reported in this work with respect to the radius of the outer shell, $R_{\text{o}}$, and the timestep width, $\Delta t$, is illustrated in figure~\ref{fig:validn_app}, for a fixed value of $\Pen=7$. In figure~\ref{fig:validn_app}~(a), the transient approach of an unforced autophoretic particle's self-propulsion speed to its steady value, calculated using a computational domain with two different values for $R_{\text{o}}$ is plotted as a function of time. While the region of transient growth is different for the two cases, they approach the same steady-state value, thus establishing convergence with respect to the outer shell radius. In figure~\ref{fig:validn_app}~(b), the transient evolution of the self-propulsion speed of a forced autophoretic particle is plotted as a function of time, for three different values of the timestep width. The agreement between the three curves establish the timestep-width convergence of the simulations.

\bibliography{ms}

\end{document}